# Power of Sine Hamiltonian Operator for Estimating the Eigenstate Energies on Quantum Computers

Qing-Xing Xie, Yi Song[*], Yan Zhao[*]

*The Institute of Technological Sciences, Wuhan University, Wuhan 430072, People's Republic of China.*

[*]Correspondence to: yi.song@whu.edu.cn (Yi Song); yan2000@whut.edu.cn (Yan Zhao)

## ABSTRACT

Quantum computers have been shown to have tremendous potential in solving difficult problems in quantum chemistry. In this paper, we propose a new classical-quantum hybrid method, named as power of sine Hamiltonian operator (PSHO), to evaluate the eigenvalues of a given Hamiltonian ($\hat{H}$). In PSHO, for any reference state $|\varphi_0\rangle$, the normalized energy of the $\sin^n(\hat{H}\tau)|\varphi_0\rangle$ state can be determined. With the increase of the power, the initial reference state can converge to the eigenstate with the largest $|\sin(E_i\tau)|$ value in the coefficients of the expansion of $|\varphi_0\rangle$, and the normalized energy of the $\sin^n(\hat{H}\tau)|\varphi_0\rangle$ state converges to $E_i$. The ground and excited state energies of a Hamiltonian can be determined by taking different $\tau$ values. The performance of the PSHO method is demonstrated by numerical calculations of the $H_4$ and LiH molecules. Compared with the current popular variational quantum eigensolver (VQE) method, PSHO does not need to design the ansatz circuits and avoids the complex nonlinear optimization problems. PSHO has great application potential in near-term quantum devices.

# 1. Introduction

Highly accurate quantum chemistry calculations are very important for understanding the intrinsic properties of molecules and materials and for elucidating the mechanisms of chemical reactions, and play a key role in material designs, biomedicine, energy, and chemical industry.[1] The core problem in quantum chemistry is to find the energy eigenstates of a given electronic Hamiltonian.[2] Of course, not only in chemistry, solving eigenvalue-problems of large matrices also has a wide applications in science and engineering, such as electrical networks, hydrodynamics, structural mechanics, etc.

The full configuration interaction (FCI) with complete basis set (CBS) limit is an exact method to calculate ground and excited states of molecules, but the computational cost increases exponentially with the size of a molecular system, which limits the its applications to systems containing only a few atoms.[3] In order to treat large molecular systems, approximate methods have been developed to balance the accuracy and computational cost, such as coupled-cluster with single and double excitation method (CCSD),[4] fourth-order Møller-Plesset perturbation theory (MP4),[5,6] and density functional theory (DFT)[7,8]. These methods are usually accurate for single-reference state quantum chemical problems where only one electronic configuration is dominant in the wavefunction. However, these methods work poorly for the simulation of stretched bond molecules, excited states, and transition metals, due to the intrinsic multireference character of these strongly correlated systems. Despite that some progresses have been achieved in extending these methods to multireference systems, the computational cost increase sharply with the increase of the number of considered reference states.[9]

Quantum computers possess profoundly different computing mode from classical devices, and quantum computing is expected to be able to solve some computational problems which are intractable in classical computing.[10,11] Developing quantum-chemistry algorithms on quantum computing devices has attracted tremendous research interests with the evolving of quantum computing hardware. In the FCI algorithm for classical computers, the resource required to store the wavefunctions increases exponentially with the increase of the number of molecular orbitals. However, due to the superposition and entanglement characteristics of quantum bits (qubits), a wavefunction in FCI is naturally suitable to be represented by a group of qubits. Only twice the number of qubits as the molecular orbitals is needed for the representation of wavefunctions, so the quantum chemical calculations can be performed much more efficiently on quantum computers than on classical ones.

Quantum phase estimation (QPE) was the first algorithm proposed to solve the Hamiltonian eigenvalue problems on a quantum computer,[12,13] in which a multiqubit register is required to control the time evolution of the Hamiltonian. The number of needed qubits depends on the required accuracy. Running the QPE algorithm, the final quantum register state will collapse to one random eigenstate of the Hamiltonian. The collapse probability depends on the overlap between the eigenstate and the initial state. By repeatedly performing QPE routine for many iterations, all the eigenstates which have non-negligible overlaps with the initial state can be obtained. Compared with the algorithms for classical computers, the QPE algorithm is shown an exponential speedup. However, the implementation of QPE requires a large qubit quantum processor with error correction and long coherence time which is a severe challenge to current Noisy Intermediate-Scale Quantum (NISQ) devices.[14] To mitigate the high quantum resource cost of QPE method, many quantum phase based algorithms have also been proposed, which partly reduces the depth of quantum circuits.[15–18]

Recently, variational quantum eigensolver (VQE), an algorithm suitable for NISQ devices, has attracted tremendous attention.[19,20] As a hybrid method, the VQE computational workflow is running on a infrastructure consisting of classical and quantum computers. A quantum processor is employed to prepare a trial state by a quantum circuit (ansatz), where the ansatz circuit is constructed by some parameterized quantum gates. By the Jordan-Wigner (or Bravyi-Kitaev) transformation, the Hamiltonian within the second quantization formulation can be mapped into a series of Pauli strings. The energy can be obtained by measuring the expectation value of these Pauli strings. In the VQE algorithm, the classical computer is employed to optimize the ansatz circuit parameters. Compared with the QPE method, the ansatz circuit depth in VQE is relatively shallow, which has obvious advantages in applying the algorithm on the near-term NISQ devices.[21] The original VQE algorithm is only applicable to the ground state calculation. Some progresses have been made in extending VQE to excited-state calculations, such as quantum subspace expansion (QSE)[22], variational quantum deflation (VQD)[23,24], subspace-search variational quantum eigensolver (SSVQE)[25,26] and orthogonal state reduction variational eigensolver (OSRVE)[27] methods.

The QPE method requires an error correction, whereas the VQE algorithm is somewhat robust to coherent errors.[28,29] However, there are two issues for the VQE-based methods. One is that the ansatz circuit should be designed rationally. Constructing an effective ansatz circuit is still an open problem. Many efforts have been devoted to the development of ansatz circuits, such as the unitary coupled cluster with single and double excitations (UCCSD)[30], qubit coupled cluster (QCC)[31–34], hardware-efficient ansatz[35], adaptive-ansatz[36–38] etc. These ansatzes can be classified into two categories. One is based on the theoretical methods in Quantum Chemistry, such as UCCSD. The other is the heuristic method, which does not rely on chemical intuition and similar to deep neural networks. However, the heuristic ansatz is suffered by the barren plateaus problem, which poses a great challenge to the optimization of quantum circuit parameters.[39] It is difficult to guarantee that the state of interest is in the representation space of the ansatz circuit. If the eigenstate is beyond the ansatz circuit representation space, it is impossible to obtain the correct results. The other drawback of VQE arises from the optimization of circuit parameters, and it is very difficult to obtain the global optimal solution for a multi-parameter, non-linear, and non-convex optimization problem. The ground state calculation requires finding the global minimum. Therefore, even a reasonable ansatz circuit is designed, the optimization method also affects the final result accuracy. Moreover, random fluctuation errors are inevitably introduced into sampling, which will further affect the performance of the gradient-based optimization algorithms.[40,41]

In addition to the QPE and VQE methods, nonunitary operation methods have attracted a widespread attention.[42–44] A typical nonunitary method is the imaginary time evolution (ITE) method, which acts the $e^{-\beta \hat{H}}$ operator on an initial state $|\varphi_0\rangle$. ITE leads the state evolve into the eigenstate $|\Psi_0\rangle$ at the evolution time limit ($\beta \to \infty$), where $|\Psi_0\rangle$ has the lowest eigenvalue of the Hamiltonian $\hat{H}$ and the overlap between $|\varphi_0\rangle$ and $|\Psi_0\rangle$ state is non-negligible. The computational complexity of ITE is prohibitively high and exponentially dependent on the number of molecular orbitals on the classical computer. Therefore, it is of great value to explore the accelerated implementation of the ITE algorithm on quantum computers. However, all qubit gates in quantum computers are unitary, which means that the nonunitary ITE method cannot be performed directly. There are three ways to indirectly implement ITE on quantum computer: variational ITE (VITE)[45,46], quantum ITE (QITE)[47,48] and probabilistic ITE (PITE)[49].

In VITE, similar with VQE, an ansatz circuit is required to prepare the trial state. The effect of ITE operator on initial state is reflected by the circuit parameters updating, in which a McLachlan's variational equation should be solved on classical computers. The essence of VITE is within the VQE algorithm framework, thus still suffered by the ansatz design problem.[50] The ansatz circuit must be selected appropriately to ensure that the evolved state can be covered by the manifold of ansatz circuit. In QITE, each ITE segment $e^{-\tau \hat{h}_i}$ is replaced by a unitary evolution operator ($e^{-i\tau \hat{A}_i}$), where $\hat{A}_i$ is a linear combination of a series of Pauli strings. A linear equation can be determined such that the unitarily evolved state approximates the exactly normalized nonunitary evolved state. The coefficients of Pauli tensors in $\hat{A}_i$ can be determined by solving the linear equation on classical computer. QITE is considered as an efficient approach to evaluate the ground state energy because it does not have the two weaknesses in the VQE-based methods and does not require deep circuits. However, the dimension of the linear equation grows exponentially with the number of qubits. In order to resolve the exponential explosion problem, the local approximation of the state is used to limit the number of Pauli strings, which would reduce the accuracy of the QITE method.[51–54] In PITE, the measured gates are used to implement nonunitary operations. Apply unitary operations on the initial state, and then perform measurements. The state collapse to the desired state with a certain probability. A complete PITE process should be divided into multiple subroutines. In each subroutine, the state should collapse to the desired state successfully, which means that the probability of successfully performing PITE decreases exponentially with the number of subroutines.

In this paper, we propose a new nonunitary operation quantum method, called power of sine Hamiltonian operator (PSHO), to evaluate the ground and excited states energy of a quantum many-body systems. For a given Hamiltonian ($\hat{H}$) and a reference state $|\varphi_0\rangle$, the normalized energy of the $\sin^n(\hat{H}\tau)|\varphi_0\rangle$ state can be determined on quantum device. The normalized $\sin^n(\hat{H}\tau)|\varphi_0\rangle$ state will converge to the eigenstate state with the maximum eigenvalue ($|\sin(E_i\tau)|$) of the $|\sin(\hat{H}\tau)|$ operator. Therefore, various eigenstates energy can be obtained by taking different $\tau$ values. Similar with the ITE methods, the implementation of PSHO also needs to overcome the nonunitary problem. It is difficult to transform the $\sin(\hat{H}\tau)$ operator into an equivalent quantum circuit. In this paper, we skillfully avoid this problem: the normalized energy of $\sin^n(\hat{H}\tau)|\varphi_0\rangle$ state is calculated without preparing the state. The method is based on the equation: $\sin(\hat{H}\tau) = \frac{i}{2}(e^{-i\hat{H}\tau} - e^{i\hat{H}\tau})$. For the $\sin^n(\hat{H}\tau)|\varphi_0\rangle$ state, the unnormalized energy can be expressed as a polynomial expansion of $e^{i\hat{H}\tau}$ with different $\tau$ values, where the expansion coefficient can be determined by the binomial theorem. We design a new quantum circuit which can be used to evaluate each term in the expansion. The unnormalized energy can be determined by substituting all of estimated values into the expansion. The normalization coefficient can be obtained by the same way, and the normalized energy can be obtained by dividing the two quantities. In addition, the normalized energy can be calculated with an efficient and low-cost route, in which only the normalization coefficients are required without calculating the unnormalized energies. With this algorithm, the sampling numbers in the quantum circuit are significantly reduced.

The rest of the paper is organized as follows. In section 2, we introduce the PSHO algorithm, including the main formulas and the technical details. The corresponding quantum circuits are proposed in this section. In section 3, we give the numerical simulation results of the $H_4$ and LiH molecules to validate our algorithm. The simulation results show that the proposed PSHO method can accurately calculate the ground and lower-order excited states. Section 4 summarizes the results and concludes the paper.

## 2. Theory and Methodology

In this section, we first discuss the PSHO theory, and propose two different implementations. The first one is a direct approach, and the normalized $\sin^n(\widehat{H}\tau)|\varphi_0\rangle$ state is directly prepared on a quantum register. The other one is an indirect method, in which the normalized energy of the $\sin^n(\widehat{H}\tau)|\varphi_0\rangle$ state is calculated without preparing the corresponding quantum state. For the sake of convenience, the unnormalized $\sin^n(\widehat{H}\tau)|\varphi_0\rangle$ state is denoted as $|\varphi_n\rangle$.

### 2.1 PSHO Theory

For any reference state $|\varphi_0\rangle$, it can be expanded into the linear superposition of $|\Psi_i\rangle$: $|\varphi_0\rangle = \sum_i c_i |\Psi_i\rangle$, where $|\Psi_i\rangle$ is the $i$th eigenstate of the Hamiltonian matrix, $c_i$ is the overlap between $|\varphi_0\rangle$ and $|\Psi_i\rangle$. Thus the $|\varphi_n\rangle$ state can be written as:

$$|\varphi_n\rangle = \sin^n(\widehat{H}\tau)|\varphi_0\rangle = \sum_i c_i \sin^n(E_i\tau)|\Psi_i\rangle \qquad (1)$$

where the energy (eigenvalue) of the $|\Psi_i\rangle$ state is denoted as $E_i$. With the increase of the power $n$, the contribution of the eigenstate $|\Psi_i\rangle$ which gives the maximum eigenvalue for the $|\sin(\widehat{H}\tau)|$ operator in $|\varphi_0\rangle$ state will increase, and the normalized $|\varphi_n\rangle = \sin^n(\widehat{H}\tau)|\varphi_0\rangle$ state will converge to the $|\Psi_i\rangle$ state. Therefore, it is necessary to explore how the $\tau$ value affect the final converged state. For the convenience of analysis, $\tau$ is limited to be greater than 0. When the value of $\tau$ satisfy $|E_0\tau| \leq \pi/2$, as can be seen in Figure 1(a), the ground state gives the maximum eigenvalue of $|\sin(E_0\tau)|$ for the $|\sin(\widehat{H}\tau)|$ operator. The relative fraction of the ground state $|\Psi_0\rangle$ in $|\varphi_n\rangle$ is exponentially increased with $n$, and the normalized $|\varphi_n\rangle$ state will converge to the ground state. The greater the difference between $\sin(E_0\tau)$ and other $\sin(E_{i\neq 0}\tau)$ values, the faster the fraction of the excited states in $|\varphi_n\rangle$ decays. It is worth noting that here we have assumed that the ground energy is negative and has the highest absolute value. This is true for most molecular systems. When the absolute value of the highest eigenvalue is greater than that of the ground state energy, the above analysis is incorrect. However, even if this is the case, it can be circumvented easily by adding a negative constant term to the Hamiltonian. The negative constant term can shift all eigenstates towards the negative direction. For the convenience of discussion, we assume that all eigenvalues are negative in the following analysis. When $|E_0\tau|$ slightly exceeds $\pi/2$, as can be seen in Figure 1(b), the ground state still gives the maximum eigenvalue for the $|\sin(\hat{H}\tau)|$ operator, but the difference of the $|\sin(\hat{H}\tau)|$ eigenvalues between the ground state and the excited state is narrowed with the increase of $\tau$, and the convergence rate will be slowed down. When the eigenvalue for $|\sin(\hat{H}\tau)|$ of the first excited state exceeds that of the ground state and is the maximum of $\sin(E_{i\neq 0}\tau)$, supposed that the overlap between the $|\varphi_0\rangle$ and $|\Psi_1\rangle$ states is non-zero, the normalized $|\varphi_n\rangle$ state will converge to the first excited state. As can be seen in Figure 1(c), when $|E_0\tau|$ is in the range of $\pi/2$ to $\pi$, with the gradual increase of $\tau$, the final state will converge to each excited state in an order. When $|E_0\tau|$ is greater than $\pi$, as shown in Figure 1(d), $|\sin(E_0\tau)|$ is increasing with $\tau$, and the converged state of $|\varphi_n\rangle$ may no longer be in an order.

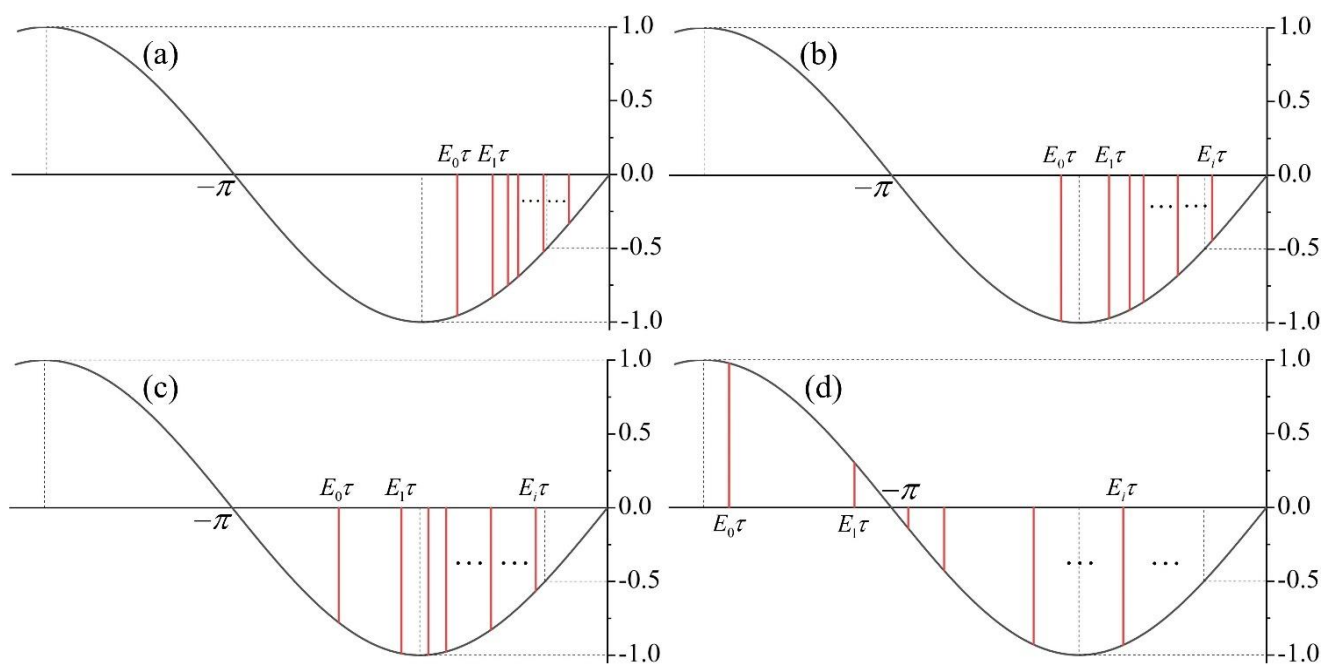

Figure 1. Schematic diagram of different $\tau$ values affecting the converged states. Y-axis represents the eigenvalue of the $\sin(\hat{H}\tau)$ operator for each eigenstate (red vertical lines). X-axis represents the $E_i\tau$ value with different $\tau$ values.

The converged energy can be obtained from the normalized and converged $|\varphi_n\rangle$ state. Take the $|E_0\tau| \leq \pi/2$ situation as an example. According to the above analysis, we have:

$$\lim_{n\to\infty} |\varphi_n\rangle = \lim_{n\to\infty} \sum_i c_i \sin^n(E_i\tau) |\Psi_i\rangle \approx c_0 \sin^n(E_0\tau) |\Psi_0\rangle \tag{2a}$$

$$\lim_{n\to\infty} E(|\varphi_n\rangle) = \lim_{n\to\infty} \frac{\langle\varphi_n|\hat{H}|\varphi_n\rangle}{\langle\varphi_n|\varphi_n\rangle} = E_0 \tag{2b}$$

Define $Q_n$ as the ratio of the normalization coefficient of $|\varphi_n\rangle$ to that of $|\varphi_{n-1}\rangle$, we have:

$$\lim_{n\to\infty} Q_n = \lim_{n\to\infty} \frac{\langle\varphi_n|\varphi_n\rangle}{\langle\varphi_{n-1}|\varphi_{n-1}\rangle} = \frac{|c_0|^2 \sin^{2n}(E_0\tau)}{|c_0|^2 \sin^{2n-2}(E_0\tau)} = \sin^2(E_0\tau) \tag{3}$$

As *n* increases, the $Q_n$ will converges to a fixed value, and the ground state energy can be calculated by:

$$E_0 = \lim_{n\to\infty} E'_n = \lim_{n\to\infty} \frac{\arcsin\sqrt{Q_n}}{\tau} \tag{4}$$

For convenience, we name the method in Eq. 2 as the energy-based PSHO method, which evaluates the normalized energy of the converged $|\varphi_n\rangle$. The method defined in Eqs 3 and 4 is named as the normalization-coefficient based PSHO method, which only evaluates the normalization coefficients.

## 2.2 Direct Method for Implementing PSHO

The direct method for implementing PSHO is to prepare the normalized $|\varphi_n\rangle$ state on a quantum computer, and evaluate its energy by sampling. Since the $\sin(\hat{H}\tau)$ operator is nonunitary, and the general quantum circuit can only support unitary gates, we need to find a way to implement the $\sin(\hat{H}\tau)$ operator on a quantum computer. As shown in previous works to implement the ITE operator on quantum computers, a promising scheme is to embed the nonunitary operator into a large unitary operator. That is, despite that the $\sin(\hat{H}\tau)$ operator is nonunitary, we have $\sin^2(\hat{H}\tau) + \cos^2(\hat{H}\tau) = \hat{I}$, where $\hat{I}$ is an identity operator. A larger unitary operator $\Sigma$ can be constructed with an ancillary qubit.

$$\Sigma = \begin{pmatrix} \sin(\hat{H}\tau) & \cos(\hat{H}\tau) \\ \cos(\hat{H}\tau) & -\sin(\hat{H}\tau) \end{pmatrix}$$

The corresponding quantum circuit can be designed as in Figure 2(a), where $H' = \frac{\sqrt{2}}{2}\begin{pmatrix} 1 & 1 \\ 1 & -1 \end{pmatrix}$ and $S = \begin{pmatrix} 1 & 0 \\ 0 & i \end{pmatrix}$. Since $\sin(\hat{H}\tau) = \frac{i}{2}(e^{-i\hat{H}\tau} - e^{i\hat{H}\tau})$ and $\cos(\hat{H}\tau) = \frac{1}{2}(e^{-i\hat{H}\tau} + e^{i\hat{H}\tau})$, two controlled-evolution gates with the opposite evolution time are used to build the $\sin(\hat{H}\tau)$ and $\cos(\hat{H}\tau)$ operators in Σ. The hollow(solid) circle in the controlled-evolution gate indicates that the evolution operator is acted only if the control qubit is 0(1). The ancillary qubit is initialized as 0, and we have:

$$|T_1\rangle = \frac{1+i}{2}|0\rangle \otimes |\varphi_0\rangle + \frac{1-i}{2}|1\rangle \otimes |\varphi_0\rangle \tag{5a}$$

$$|T_2\rangle = \frac{1+i}{2}|0\rangle \otimes e^{-iH\tau}|\varphi_0\rangle + \frac{1-i}{2}|1\rangle \otimes e^{iH\tau}|\varphi_0\rangle \tag{5b}$$

$$|T_3\rangle = |0\rangle \otimes \sin(\hat{H}\tau)|\varphi_0\rangle + |1\rangle \otimes \cos(\hat{H}\tau)|\varphi_0\rangle = \begin{pmatrix} \sin(\hat{H}\tau)|\varphi_0\rangle \\ \cos(\hat{H}\tau)|\varphi_0\rangle \end{pmatrix} \tag{5c}$$

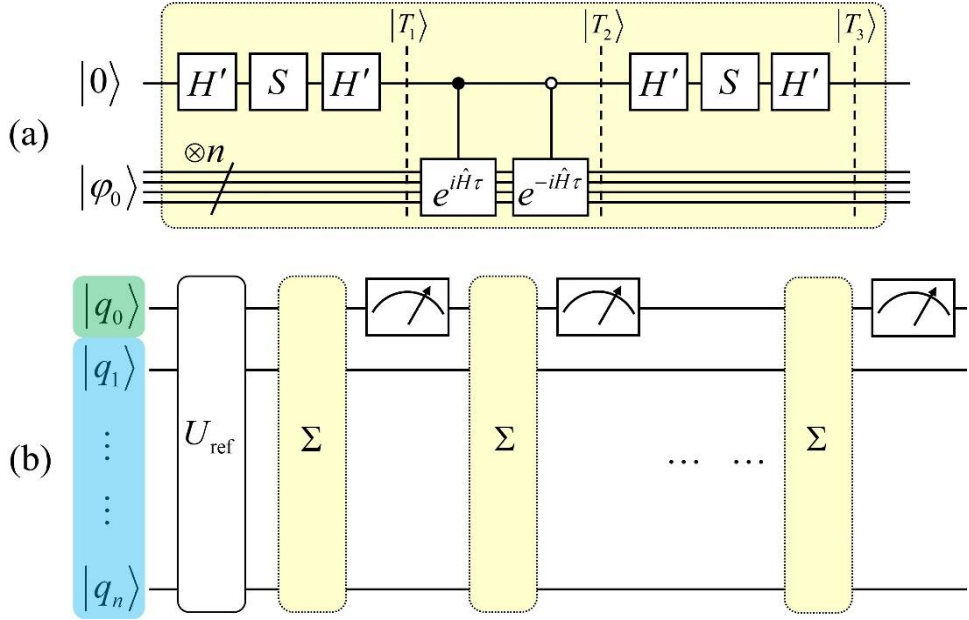

Figure 2. (a) Quantum circuit for the Σ operator. (b) Quantum circuit architecture for preparing the normalized $\sin^n(\hat{H}\tau)|\varphi_0\rangle$ state. The $q_0$ (in the green box) is used as the ancillary qubit, and the $q_1 \cdots q_n$ (in the cyan box) are register qubits. This color code is also applied in Figures 3.

After the $|T_3\rangle$ state is prepared, a measuring gate on the ancillary qubit is used to extract the target state. If the auxiliary qubit collapses to the 0 state, the register qubits will collapse to the normalized $\sin(\hat{H}\tau)|\varphi_0\rangle$ state. The quantum circuit architecture for performing the $\sin^n(\hat{H}\tau)$ operator can be seen in Figure 2(b). The reference state $|\varphi_0\rangle$ is prepared by the $U_{\text{ref}}$ circuit. There are many ways to prepare the reference state. The simplest choice is the HF state, and only $N_e$ X gates ($X = \begin{pmatrix} 0 & 1 \\ 1 & 0 \end{pmatrix}$) are needed to construct the $U_{\text{ref}}$ circuit, where $N_e$ is the number of electrons in the Hamiltonian. These X gates act on the $N_e$ qubits produce the lowest energy spatial-spin orbitals. Furthermore, the VQE calculation can be implemented to obtain a set of optimized parameterized quantum circuits. Even though the accuracy of the VQE result may not reach the required accuracy, the VQE-state is closer to the ground state than the HF state. Using VQE state as the reference state in the PSHO algorithm, the power required to converge to the ground state may be less than that of the HF state. In this paper, unless otherwise specified, the HF state is used as the reference state by default. Then repeat $\Sigma$ and measuring circuits for $n$ times. If auxiliary qubit collapses to the 0 state in each sampling, the normalized $|\varphi_n\rangle$ state is successfully prepared in the register qubits, and the energy can be evaluated accordingly.

The probability of successfully preparing the normalized $|\varphi_n\rangle$ state can be calculated as follows. For $n = 1$, with the initial state is $|\varphi_0\rangle$, the probability for sampling the 0 state on the auxiliary qubit is $\langle\varphi_0|\sin^2(\widehat{H}\tau)|\varphi_0\rangle$. After the successful collapse, the register qubit state changes into $d_1|\varphi_1\rangle$, where $d_1 = \langle\varphi_0|\sin^2(\widehat{H}\tau)|\varphi_0\rangle^{-\frac{1}{2}}$ is the normalization coefficient. Acting the $\Sigma$ operator and measuring circuits again, the probability for sampling the 0 state is $|d_1|^2\langle\varphi_1|\sin^2(\widehat{H}\tau)|\varphi_1\rangle = |d_1|^2\langle\varphi_0|\sin^4(\widehat{H}\tau)|\varphi_0\rangle$. Similarly, for the $i$th time, the register qubits state changes into $d_{i-1}|\varphi_{i-1}\rangle$, where $d_{i-1} = \langle\varphi_0|\sin^{2(i-1)}(\widehat{H}\tau)|\varphi_0\rangle^{-\frac{1}{2}}$, and the success probability is $|d_{i-1}|^2\langle\varphi_{i-1}|\sin^2(\widehat{H}\tau)|\varphi_{i-1}\rangle = |d_{i-1}|^2\langle\varphi_0|\sin^{2i}(\widehat{H}\tau)|\varphi_0\rangle$. In order to obtain the normalized $|\varphi_n\rangle$ state, each sampling in auxiliary qubit must be 0 state. Therefore, the probability of constantly collapsing to 0 state $n$ times is the product of the probability of collapse to the zero state in each step, and is given by:

$$\begin{aligned} P_{00\cdots0} &= \langle\varphi_0|\sin^2(\widehat{H}\tau)|\varphi_0\rangle \cdot |d_1|^2\langle\varphi_1|\sin^2(\widehat{H}\tau)|\varphi_1\rangle \cdots |d_n|^2\langle\varphi_{n-1}|\sin^2(\widehat{H}\tau)|\varphi_{n-1}\rangle \\ &= \langle\varphi_0|\sin^2(\widehat{H}\tau)|\varphi_0\rangle \cdot \frac{\langle\varphi_0|\sin^4(\widehat{H}\tau)|\varphi_0\rangle}{\langle\varphi_0|\sin^2(\widehat{H}\tau)|\varphi_0\rangle} \cdots \frac{\langle\varphi_0|\sin^{2n}(\widehat{H}\tau)|\varphi_0\rangle}{\langle\varphi_0|\sin^{2(n-1)}(\widehat{H}\tau)|\varphi_0\rangle} \\ &= \langle\varphi_0|\sin^{2n}(\widehat{H}\tau)|\varphi_0\rangle \end{aligned} \quad (6)$$

It is obvious that the probability for preparing the desired state in the register qubits decreases exponentially with the increase of $n$, which makes this direct scheme not practical.

**2.3 Indirect Method for Implementing PSHO**

The normalized energy of the $|\varphi_n\rangle$ state is:

$$E(|\varphi_n\rangle) = \frac{\langle\varphi_n|\widehat{H}|\varphi_n\rangle}{\langle\varphi_n|\varphi_n\rangle} = \frac{\langle\varphi_0|\widehat{H}\sin^{2n}(\widehat{H}\tau)|\varphi_0\rangle}{\langle\varphi_0|\sin^{2n}(\widehat{H}\tau)|\varphi_0\rangle} \quad (7)$$

Substitute $\sin(\widehat{H}\tau) = \frac{i}{2}(e^{-i\widehat{H}\tau} - e^{i\widehat{H}\tau})$ into Eq. 7, we have:

$$E(|\varphi_n\rangle) = \frac{\left\langle \varphi_0 \left| \left(\frac{i}{2}\right)^{2n} \cdot \hat{H} \cdot \left(e^{-i\hat{H}\tau} - e^{i\hat{H}\tau}\right)^{2n} \right| \varphi_0 \right\rangle}{\left\langle \varphi_0 \left| \left(\frac{i}{2}\right)^{2n} \cdot \left(e^{-i\hat{H}\tau} - e^{i\hat{H}\tau}\right)^{2n} \right| \varphi_0 \right\rangle} = \frac{\sum_{k=0}^{2n}(-1)^{2n-k} \cdot C_{2n}^k \cdot \langle \varphi_0 | \hat{H} e^{i\hat{H}\tau \cdot (2n-2k)} | \varphi_0 \rangle}{\sum_{k=0}^{2n}(-1)^{2n-k} \cdot C_{2n}^k \cdot \langle \varphi_0 | e^{i\hat{H}\tau \cdot (2n-2k)} | \varphi_0 \rangle} \quad (8)$$

For the sake of convenience, the numerator of Eq. 8 is denoted as $A_n$, while the denominator is $B_n$. Here we explore the feasibility of evaluating each term of $A_n$ and $B_n$ individually on quantum device. For the term in the denominator of Eq. 8, due to $C_{2n}^k = C_{2n}^{2n-k}$, the $k$th term has the same coefficient as the $2n$-$k$th term and can be paired together. We have:

$$\begin{aligned}
&(-1)^{2n-k} \cdot C_{2n}^k \cdot \langle \varphi_0 | e^{i\hat{H}\tau \cdot (2n-2k)} | \varphi_0 \rangle + (-1)^{2n-(2n-k)} \cdot C_{2n}^{2n-k} \cdot \langle \varphi_0 | e^{i\hat{H}\tau \cdot (2n-2(2n-k))} | \varphi_0 \rangle \\
&= (-1)^k \cdot C_{2n}^k \cdot \langle \varphi_0 | e^{i\hat{H}\tau \cdot (2n-2k)} + e^{i\hat{H}\tau \cdot (2k-2n)} | \varphi_0 \rangle \\
&= (-1)^k \cdot 2C_{2n}^k \cdot \langle \varphi_0 | \cos[\hat{H}\tau \cdot 2(n-k)] | \varphi_0 \rangle
\end{aligned} \quad (9)$$

Substitute Eq. 9 into $B_n$ (denominator of Eq. 8), we have:

$$B_n = \sum_{k=0}^{n-1}(-1)^k \cdot 2C_{2n}^k \cdot \langle \varphi_0 | \cos[\hat{H}\tau \cdot 2(n-k)] | \varphi_0 \rangle + C_{2n}^n \cdot (-1)^n \quad (10)$$

Using the same derivation, $A_n$ (numerator of Eq. 8) can be written as:

$$A_n = \sum_{k=0}^{n-1}(-1)^k \cdot 2C_{2n}^k \cdot \langle \varphi_0 | \hat{H} \cos[\hat{H}\tau \cdot 2(n-k)] | \varphi_0 \rangle + C_{2n}^n \cdot (-1)^n \langle \varphi_0 | \hat{H} | \varphi_0 \rangle \quad (11)$$

Therefore, the problem is transformed into the estimations of the $\langle \varphi_0 | \cos[\hat{H}\tau \cdot 2(n-k)] | \varphi_0 \rangle$ and $\langle \varphi_0 | \hat{H} \cos[\hat{H}\tau \cdot 2(n-k)] | \varphi_0 \rangle$ terms on quantum computers. Using $\Sigma$ operator circuit in Figure 2(a) and setting evolution time as $\tau(n-k)$, the output $|T_3\rangle$ state is:

$$|T_3\rangle = |0\rangle \otimes \sin[\hat{H}\tau(n-k)]|\varphi_0\rangle + |1\rangle \otimes \cos[\hat{H}\tau(n-k)]|\varphi_0\rangle = \begin{pmatrix} \sin[\hat{H}\tau(n-k)]|\varphi_0\rangle \\ \cos[\hat{H}\tau(n-k)]|\varphi_0\rangle \end{pmatrix} \quad (12)$$

For the $|T_3\rangle$ state, the expectation value of $-\hat{Z} \otimes \hat{I}$ and $-\hat{Z} \otimes \hat{H}$ are:

$$\begin{aligned}
\langle T_3 | -\hat{Z} \otimes \hat{I} | T_3 \rangle &= \langle \varphi_0 | \cos^2[\hat{H}\tau(n-k)] | \varphi_0 \rangle - \langle \varphi_0 | \sin^2[\hat{H}\tau(n-k)] | \varphi_0 \rangle \\
&= \langle \varphi_0 | \cos[\hat{H}\tau \cdot 2(n-k)] | \varphi_0 \rangle
\end{aligned} \quad (13a)$$

$$\begin{aligned}
\langle T_3 | -\hat{Z} \otimes \hat{H} | T_3 \rangle &= \langle \varphi_0 | \hat{H} \cdot \cos^2[\hat{H}\tau(n-k)] | \varphi_0 \rangle - \langle \varphi_0 | \hat{H} \cdot \sin^2[\hat{H}\tau(n-k)] | \varphi_0 \rangle \\
&= \langle \varphi_0 | \hat{H} \cdot \cos[\hat{H}\tau \cdot 2(n-k)] | \varphi_0 \rangle
\end{aligned} \quad (13b)$$

They are identical to the $k$th term of $A_n$ and $B_n$, respectively. Therefore, the normalized energy of the $|\varphi_n\rangle$ state can be determined by the following scheme. Take $k$ value from 0 to $n-1$ and substitute it into the quantum circuit diagram of Figure 3, then measure the expectation value of $-\hat{Z} \otimes \hat{H}$ and $-\hat{Z} \otimes \hat{I}$ for the output state, so that the $\langle \varphi_0 | \hat{H} \cos[\hat{H}\tau \cdot 2(n-k)] | \varphi_0 \rangle$ and $\langle \varphi_0 | \cos[\hat{H}\tau \cdot 2(n-k)] | \varphi_0 \rangle$ terms can be evaluated. The last term in $A_n$ is $C_{2n}^n \cdot (-1)^n \langle \varphi_0 | \hat{H} | \varphi_0 \rangle$,

in which $\langle\varphi_0|\hat{H}|\varphi_0\rangle$ can be calculated on classical computers easily. Substituting all of them into Eq. 10 and Eq. 11, $A_n$ and $B_n$ can be determined.

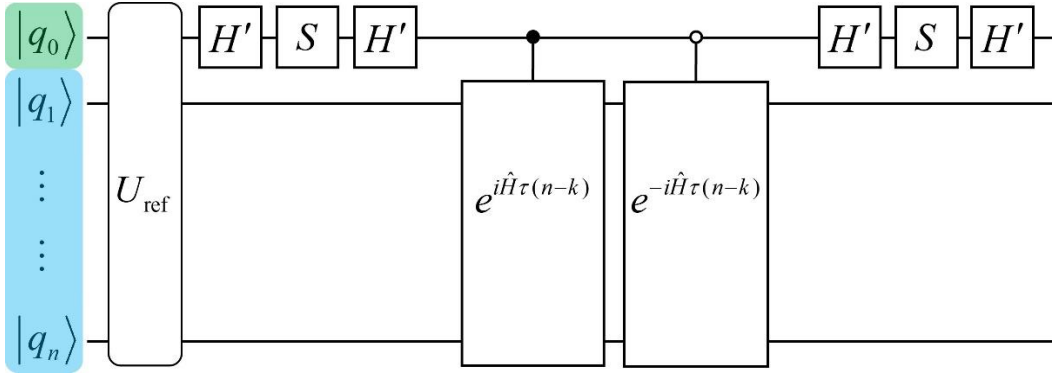

Figure 3. Quantum circuit architecture for calculating $\langle\varphi_0|\hat{H}\cos[\hat{H}\tau\cdot 2(n-k)]|\varphi_0\rangle$ and $\langle\varphi_0|\cos[\hat{H}\tau\cdot 2(n-k)]|\varphi_0\rangle$ terms. For the former, the expectation operator should be set as $-\hat{Z}\otimes\hat{H}$, for the latter that should be $-\hat{Z}\otimes\hat{I}$.

Compared with the direct method in Section 2.2, the indirect method described in this section does not suffered from the problem that the sampling number increases exponentially with the power. Multiple sampling is also required in the indirect method, but the sampling number is only related to the number of Pauli strings in the Hamiltonian and the measuring accuracy, not the power. Therefore, the indirect method has more practical application potential than the direct method. In the numerical simulation of Section 3, we will use the indirect method to perform the PSHO calculations. Before that, it is necessary to analyze the systematic errors in the final result, which arise from the Suzuki-Trotter decomposition of the time-evolution operators and the sampling accuracy. The former can be reduced by using higher order Suzuki-Trotter decomposition, while the latter one can be reduced by increasing the number of samples. However, the systematic error cannot be completely eliminated. Even when the error is small, the $C_{2n}^k$ coefficient increases factorially with $n$, the influence of the error on the final result will be dramatically amplified with the increase of the power. For the ease of the error analysis, the total error in evaluating $\langle\varphi_0|\hat{H}\cos[\hat{H}\tau\cdot 2(n-k)]|\varphi_0\rangle$ and $\langle\varphi_0|\cos[\hat{H}\tau\cdot 2(n-k)]|\varphi_0\rangle$ are denoted as $\Delta_k$ and $\delta_k$, respectively. The accurate $A_n'$ and $B_n'$ can be expressed as:

$$A_n' = \sum_{k=0}^{n-1}(-1)^k\cdot 2C_{2n}^k\cdot\langle\varphi_0|\hat{H}\cos[\hat{H}\tau\cdot 2(n-k)]|\varphi_0\rangle + (-1)^n C_{2n}^n\langle\varphi_0|\hat{H}|\varphi_0\rangle + \sum_{k=0}^{n-1}(-1)^k\cdot 2C_{2n}^k\Delta_k \quad (14a)$$

$$B_n' = \sum_{k=0}^{n-1}(-1)^k\cdot 2C_{2n}^k\cdot\langle\varphi_0|\cos[\hat{H}\tau\cdot 2(n-k)]|\varphi_0\rangle + (-1)^n C_{2n}^n + \sum_{k=0}^{n-1}(-1)^k\cdot 2C_{2n}^k\delta_k \quad (14b)$$

According to Eq. 1, Eq. 8 and Eq. 9, we have:

$$A'_n = (-1)^n \sum_j 2^{2n}|c_j|^2 E_j \sin^{2n}(E_j\tau) + \sum_{k=0}^{n-1}(-1)^k \cdot 2C_{2n}^k \Delta_k \qquad (15a)$$

$$B'_n = (-1)^n \sum_j 2^{2n}|c_j|^2 \sin^{2n}(E_j\tau) + \sum_{k=0}^{n-1}(-1)^k \cdot 2C_{2n}^k \delta_k \qquad (15b)$$

The first term on the right-hand side of Eq. 15 is denoted as the information term, while the second is denoted as the error term. The information term itself can be accurately calculated, but the introduction of error term cause the deviation from the accurate results. Only if the information term is much larger than the error term the result will not have obvious deviation. The upper limit of the error term in $B'_n$ is:

$$\begin{aligned}\sum_{k=0}^{n-1}(-1)^k \cdot 2C_{2n}^k \delta_k &\leq |\delta_{max}| \cdot \sum_{k=0}^{n-1} 2C_{2n}^k = |\delta_{max}| \cdot \left(\sum_{k=0}^{n-1} C_{2n}^k + \sum_{k=0}^{n-1} C_{2n}^{2n-k}\right) \\ &= |\delta_{max}| \cdot \left(\sum_{k=0}^{2n} C_{2n}^k - C_{2n}^n\right) = |\delta_{max}| \cdot (2^{2n} - C_{2n}^n)\end{aligned} \qquad (16)$$

where the $|\delta_{max}|$ is denoted as the absolute maximum error in $\{\delta_k | k = 0,1,\cdots n-1\}$, which can be far less than 1 by increasing the order of Suzuki-Trotter decomposition and the number of samples. According to Eq.15a and Eq.16, when the maximum of $|\sin(E_i\tau)|$ is close to 1, the scaling factors of the information and error terms are almost identical, both of which are approximately $2^{2n}$. Strictly speaking, since $C_{2n}^n$ should to be subtracted from $2^{2n}$, the scaling factor of the error term will be slightly less than $2^{2n}$. Similarly, the upper limit of the error term in $A'_n$ is $|\Delta_{max}| \cdot (2^{2n} - C_{2n}^n)$, and the scaling factors of the information terms in $A'_n$ and $B'_n$ are equal. Therefore, in order to ensure the reliability of the PSHO calculations, the setting of $\tau$ should ensure that the maximum $|\sin(E_i\tau)|$ is close to 1. Only with this choice, the information term is always far greater than the error term. When the $|\sin(E_{max}\tau)|$ is obviously less than 1, the scale of the error term will gradually approach and even exceed the information term with the increase of power. The smaller the $|\sin(E_{max}\tau)|$ is, the faster the error term exceeds the information term.

It is worth mentioning that there are some other recent works which is also nonunitary operation methods, such as the cooling method[55,56], the filtering method[57,58], and the quantum power method[59]. By using the Fourier transformation, the cooling method is to expand the imaginary time evolution operator into the integral form of the time evolution operator over time. The time evolution operator can be implemented on quantum devices easily. However, the time in the integral expansion is continuous, which means that the time evolution should be performed separately at all times. Therefore, it should be executed many times. In order to avoid this problem, the idea of quantum Monte Carlo is employed. Different evolve-times are selected by finite sampling, and the imaginary time evolution is approximated by the linear combination of these different time evolutions. The idea of quantum Monte Carlo is also used in the filtering method, but the executed nonunitary operator is not the imaginary time evolution operator, but the filter function of Hamiltonian operator. This filtering operator can filter out energies outside the desired energy interval. All of these methods are different from the PSHO method, not only in the selection of nonunitary operators, but also in the implementation of nonunitary operators on quantum devices. Furthermore, it is well known that the quantum Monte Carlo method suffer from the notorious sign problem, which makes the sampling variance exponentially increase with the system size, and the cooling and filtering methods may suffer the sign problem too.

In the quantum power method, the ground state can be evaluated by simulate the Hamiltonian power operator ($\hat{H}^n$) on a reference state. For the $\hat{H}$ operator, we have $\hat{H} = \lim_{\Delta \to 0} \frac{i}{2\Delta}\left(\hat{U}(\Delta) - \hat{U}(-\Delta)\right)$, where the $\hat{U}(\Delta)$ represents the $\Delta$-time evolution ($\hat{U}(\Delta) = e^{-i\hat{H}\Delta}$). Thus the $\hat{H}^n$ operator can be approximated by a linear combination of time evolution operators formally given by $\hat{H}^n = \left(\frac{i}{2\Delta}\right)^n \sum_{k=0}^{n} C_n^k [\hat{U}(\Delta)]^{n-2k}$, where $C_n^k$ is binomial expansion coefficient. Each $\left\langle \varphi_0 \left| [\hat{U}(\Delta)]^{n-2k} \right| \varphi_0 \right\rangle$ term can be evaluated on quantum devices easily (by Hadamard test) and the classical computer perform the multiplications and sum. It can be seen that the quantum power method shares the similar spirit of our PSHO method. Compared with the $\sin^n(\hat{H}\tau)$ operator, the $\hat{H}^n$ operator can separate different eigenstates further. In the vast majority of cases, we have $\left|\frac{E_0}{E_1}\right| > \left|\frac{\sin(E_0\tau)}{\sin(E_1\tau)}\right|$, which means that the $\hat{H}^n|\varphi_0\rangle$ state converges to the ground state faster than the $\sin^n(\hat{H}\tau)|\varphi_0\rangle$ state with power increasing. However, PSHO can calculate any eigenstates, while the quantum power method can only calculate the ground state, because $\hat{H}^n$ operator can only filter out the state with the lowest energy. In contrast, the $\sin^n(\hat{H}\tau)$ operator can filter out various eigenstates by varying the time parameter $\tau$. Moreover, the quantum power method faces a potential problem. That is, in order to ensure that the time evolution expansion approximates to the Hamiltonian operator, the time step $\Delta$ must be very small. According to the above analysis on error, when $\Delta \to 0$, the scale of the error term will gradually exceed the information term with the increase of power, and the calculation result is no longer reliable. Therefore, although the quantum power method requires a lower power than the PSHO method to achieve the same accuracy, the calculation results of the quantum power method may become inaccurate with the increase of power.

## 3. Numerical Simulation Results and Discussion

In this section, we present numerical results to validate the proposed algorithm. Before that, it is necessary to explain how to decompose the time evolution operator into executable quantum circuits and analyze the errors caused by the transformation, and this is discussed in section 3.1. In Section 3.2, we demonstrated the effect of different $\tau$ on the final results, which confirmed our analysis in Section 2.3. In Section 3.3, we calculate the ground state energy for the $H_4$ and LiH molecules by the proposed indirect PSHO methods. In Section 3.4, we calculate lower-order excited states energies for the $H_4$ molecule. In Section 3.5, we discuss the feasibility of implementing the PSHO algorithm on quantum devices based on previous theoretical analysis and numerical simulation results, and compare it with the current popular VQE method. All of the quantum simulations are implemented in MindQuantum, which is a general framework to build and simulate various quantum circuits.[60] The PySCF package is used to calculate the molecular orbitals.[61,62] The OpenFermion package is used to convert fermion operators to qubit operators by the JW transformation.[63]

### 3.1 Suzuki-Trotter Decomposition

In the indirect method for PSHO, two controlled-evolution gates with opposite evolution time are required. The Hamiltonian can be mapped into a series of Pauli strings: $\hat{H} = \hat{h}_1 + \hat{h}_2 + \cdots + \hat{h}_Z$, and the first-order ($\hat{U}_1(\tau)$) and second-order ($\hat{U}_2(\tau)$) Suzuki-Trotter decomposition can be written as:[64–66]

$$\hat{U}_1(\tau) = e^{-i\hat{h}_1\tau}e^{-i\hat{h}_2\tau}\cdots e^{-i\hat{h}_Z\tau} \tag{17a}$$

$$\hat{U}_2(\tau) = e^{-\frac{i\hat{h}_1\tau}{2}}e^{-\frac{i\hat{h}_2\tau}{2}}\cdots e^{-i\hat{h}_Z\tau}\cdots e^{-\frac{i\hat{h}_2\tau}{2}}e^{-\frac{i\hat{h}_1\tau}{2}} \tag{17b}$$

The second-order Suzuki-Trotter decomposition satisfies time inversion symmetry, that is $[\hat{U}_2(\tau)]^\dagger = \hat{U}_2(-\tau)$ and $\hat{U}_2(\tau)\hat{U}_2(-\tau) = \hat{U}_2(-\tau)\hat{U}_2(\tau) = \hat{I}$, while it is not satisfied in the first-order decomposition. Given that $\sin(\hat{H}\tau) = \frac{i}{2}(e^{-i\hat{H}\tau} - e^{i\hat{H}\tau})$, if the $\hat{U}_1(\tau)$ circuit is used to approximate the $e^{-i\hat{H}\tau}$ and $e^{i\hat{H}\tau}$ operators, the approximated $\sin(\hat{H}\tau)$ operator no longer satisfies hermiticity. But the hermiticity can still be maintained in the $\hat{U}_2(\tau)$ circuit. Therefore, in the present study, the second-order Suzuki-Trotter decomposition is used to approximate the time-evolution operators ($e^{-i\hat{H}\tau}$) in numerical simulations. According to the previous studies[65–67], the second-order Suzuki-Trotter decomposition satisfies:

$$e^{-i\hat{H}\tau} = \left(e^{-\frac{i\hat{h}_1\Delta}{2}}e^{-\frac{i\hat{h}_2\Delta}{2}}\cdots e^{-i\hat{h}_Z\Delta}\cdots e^{-\frac{i\hat{h}_2\Delta}{2}}e^{-\frac{i\hat{h}_1\Delta}{2}}\right)^{\frac{\tau}{\Delta}} + O(\tau \cdot \Delta^2) \tag{18}$$

where $\tau$ should be divisible by $\Delta$. We calculated the errors in the Suzuki-Trotter decomposition with different $\Delta$ and $\tau$ by numerical simulations. Figure 4 exhibits the deviation of $\langle\varphi_0|\hat{H}\cos(2\hat{H}\tau)|\varphi_0\rangle$ and $\langle\varphi_0|\cos(2\hat{H}\tau)|\varphi_0\rangle$ from their exact values, in which the Hamiltonian is for the H$_4$ molecular chain with the bond length of 1.0Å using the STO-3G basis set. Different $\Delta$ values are set for evaluating by quantum computers. If $\tau$ is not divisible by $\Delta$, the evolution circuit can be decomposed as:

$$\left(e^{-\frac{i\hat{h}_1\Delta}{2}}e^{-\frac{i\hat{h}_2\Delta}{2}}\cdots e^{-i\hat{h}_Z\Delta}\cdots e^{-\frac{i\hat{h}_2\Delta}{2}}e^{-\frac{i\hat{h}_1\Delta}{2}}\right)^{\lfloor\frac{\tau}{\Delta}\rfloor}\left(e^{-\frac{i\hat{h}_1\tau_0}{2}}e^{-\frac{i\hat{h}_2\tau_0}{2}}\cdots e^{-i\hat{h}_Z\tau_0}\cdots e^{-\frac{i\hat{h}_2\tau_0}{2}}e^{-\frac{i\hat{h}_1\tau_0}{2}}\right)$$

where $\lfloor\frac{\tau}{\Delta}\rfloor$ is the integer part of $\frac{\tau}{\Delta}$, and $\tau_0 = \tau - \Delta \cdot \lfloor\frac{\tau}{\Delta}\rfloor$. As can be seen in Figure 4, all curves with different $\Delta$ share a similar appearance. All deviation curves show an obvious oscillation trend, and the scale of all extreme points increase linearly with the evolution time. Different $\Delta$ values determine the slope of this linear scaling relationship. A larger value of $\Delta$ results in a more obvious increase of the deviation with the evolution time.

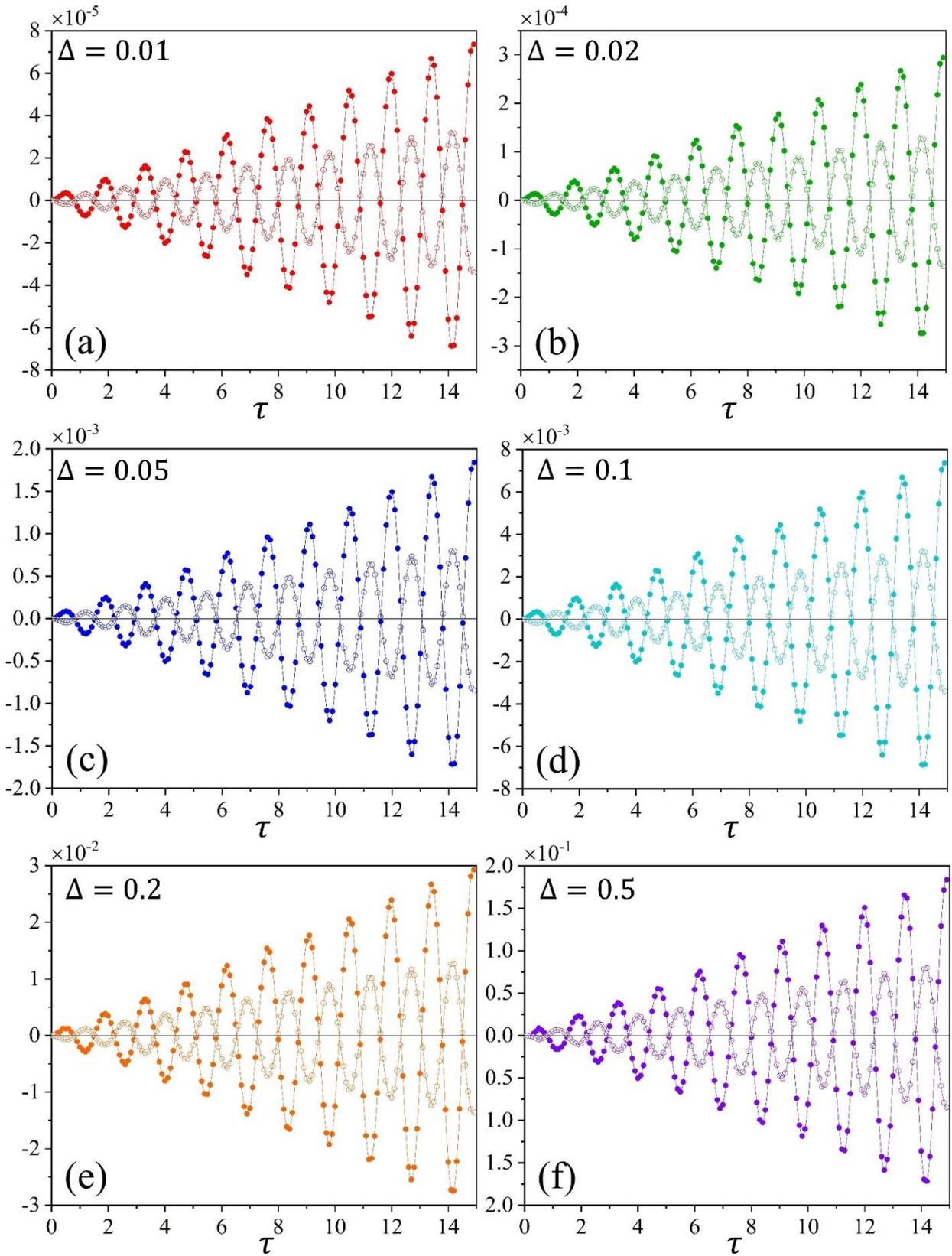

Figure 4. Deviation of evaluated $\langle\varphi_0|\hat{H}\cos(2\hat{H}\tau)|\varphi_0\rangle$ (solid circle) and $\langle\varphi_0|\cos(2\hat{H}\tau)|\varphi_0\rangle$ (hollow circle) values from their exact value. The results are for the $H_4$ molecular chain with 1.0Å bond length (STO-3G basis set). Y-axis represents deviation, and X-axis represents different $\tau$ values. The solid line in figures is obtained by cubic B-spline interpolation, which is a good visual guide. Different $\Delta$ values are used for the second-order Suzuki-Trotter decomposition: (a) $\Delta = 0.01$, (b) $\Delta = 0.02$, (c) $\Delta = 0.05$, (d) $\Delta = 0.1$, (e) $\Delta = 0.2$, (f) $\Delta = 0.5$.

In order to further explore the influence of $\Delta$ on the deviation, a quantitative analysis is given in Figure 5. As can be seen in Figure 5(a), for each deviation curve, the envelopes of these curves are linearly fitted and the slopes are calculated. Each curve has two envelope lines, namely, the upper and lower ones. The absolute values of the slopes obtained by fitting the two envelopes are almost identical. The deviation scaling factor can be defined by the average of the absolute values of the two slopes. Figure 5(b) shows the relationship between the deviation scaling factor and $\Delta$. The quadratic curve with $y = ax^2$ form is used to fit the scatter points in the figure. The fitted curves are perfectly consistent with sampling points, and the fitting error is within $2 \times 10^{-5}$. which is consistent with the conclusions in the literature.

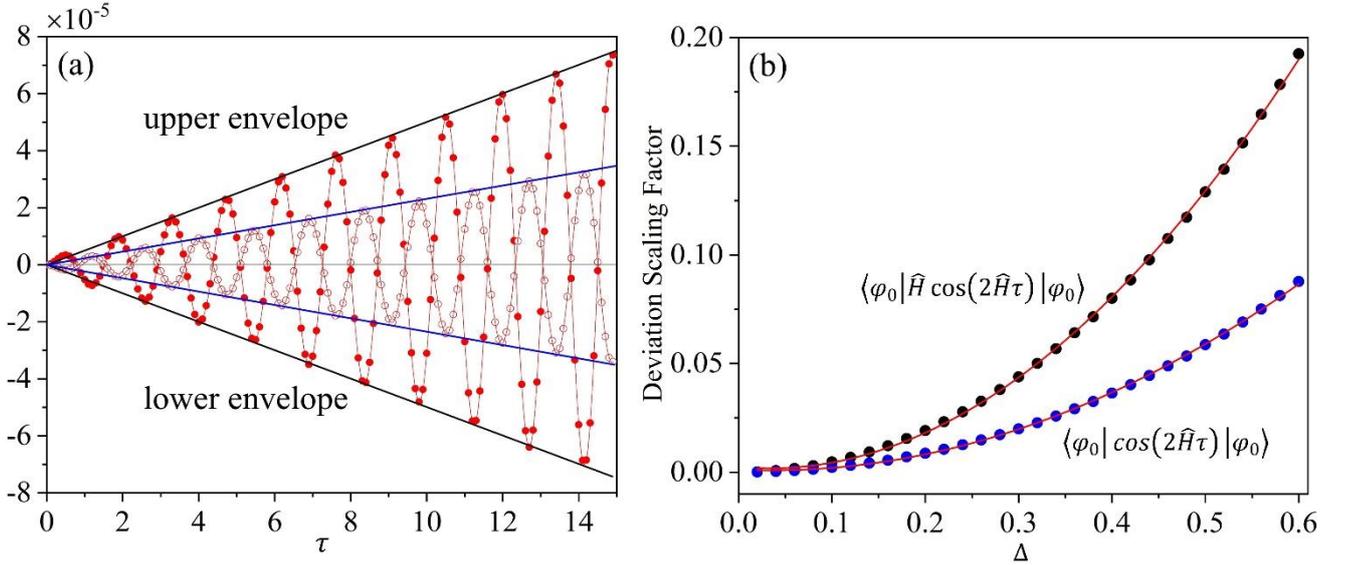

Figure 5. (a) Take Figure 4(a) as an example to illustrate the method of calculating the deviation scaling factor, where two black(blue) solid lines are obtained by linear fitting upper and lower envelopes. (b) The deviation scaling factor as a function of $\Delta$. The solid circles are the calculated result by sampling different $\Delta$ value, in which the black solid circle is the deviation scaling factor of $\langle\varphi_0|\hat{H}\cos(2\hat{H}\tau)|\varphi_0\rangle$, and the blue solid circle is for $\langle\varphi_0|\cos(2\hat{H}\tau)|\varphi_0\rangle$. The solid lines are obtained by quadratic fitting with the fitting curve is $y = ax^2$.

Now we analyze the cost in simulating a controlled-evolution circuit under a fixed deviation. Since the deviation ($\varepsilon$) can be expressed as $\varepsilon = O(\tau \cdot \Delta^2)$. The number of transformed Pauli string terms ($\hat{h}_i$) in a Hamiltonian is $O(N^4)$, where $N$ is the number of spatial-spin orbitals (which is equal to the number of qubits in Jordan-Wigner encoding). The depth of each controlled $e^{-i\hat{h}_i\Delta}$ circuit is $O(N)$, so the gate cost in simulating the controlled $e^{-i\hat{H}\Delta}$ operator is $O(N^5)$. The controlled $e^{-i\hat{H}\Delta}$ operator should be repeated $\frac{\tau}{\Delta}$ times. According to the analysis above, the total gate cost ($N_G$) in simulating the controlled $e^{-i\hat{H}\tau}$ circuit with deviation within $\varepsilon$ is:

$$N_G = \frac{\tau}{\Delta} \cdot O(N^5) = O\left(\sqrt{\frac{\tau}{\varepsilon}} \cdot N^5\right) \quad (19)$$

It is worth noting that Eq. 19 does not take into consideration of the measurement statistical error, so $N_G$ in Eq. 19 is the lower limit of the total gate cost. In the actual implementation of the PSHO method, if the precision is $\varepsilon_0$, the precision $\varepsilon$ substituted into Eq. 19 should be less than $\varepsilon_0$. Because measurement sampling will introduce some statistical error. Given a total error $\varepsilon_0$, the precision $\varepsilon$ substituted into Eq. 19 depends on the number of quantum measurements. The more measurement numbers, the smaller the measurement statistical error, which makes the maximum Trotter decomposition precision closer to $\varepsilon_0$, thus the number of quantum gates will be reduced. Therefore, the actual number of required gates to simulate the controlled $e^{-i\hat{H}\tau}$ circuit is more than (or close to) that estimated in Eq. 19.

### 3.2 Error Analysis in the Indirect Implementation of PSHO

According to the analysis in Section 2.3, when $|\sin(E_{max}\tau)|$ is close to 1, the numerical error is not significant, and the PSHO calculation is very accurate. However, if $|\sin(E_{max}\tau)|$ deviates far from 1, the error will be large, and the PSHO result is not accurate at all. It is clear that $|\sin(E_{max}\tau)|$ is directly determined by $\tau$. In this section, we demonstrated the effect of different $\tau$ values on the final results in the ground state calculation with the indirect PSHO method. The numerical simulation system is the H$_4$ molecular chain with the bond length of 1.0Å (STO-3G basis set). To better estimate the errors, for all of $\langle\varphi_0|\hat{H}\cos[\hat{H}\tau\cdot 2(n-k)]|\varphi_0\rangle$ and $\langle\varphi_0|\cos[\hat{H}\tau\cdot 2(n-k)]|\varphi_0\rangle$ values estimated on quantum computers, only six decimal places are kept. The dropped decimal part is equivalent to the error, which will affect the final results.

Figure 6 presents the ground energies for the H$_4$ molecular with the bond length of 1.0Å calculated by the indirect PSHO method. The accurate FCI/STO-3G ground state energy is $E_0 = -2.166$ hartree and we have $\left|\frac{\pi}{2E_0}\right| = 0.725$. As can be seen in Figure 6(a)(b), when $\tau$ is 0.85, the $E(|\varphi_n\rangle)$ does not converge to $E_0$. As the $\tau$ value decreases, the convergence becomes better. When $\tau$ is less than or equal to 0.6, $E(|\varphi_n\rangle)$ does not converged to a fixed value with the increase of power. In order to further explore the influence of $\tau$ on $E(|\varphi_n\rangle)$, Figure 6(c)(d) plot $E(|\varphi_n\rangle)$ vs $\tau$ for different power $n$. Obviously, with $\tau$ is in the range of 0.62 to 0.8, a plateau is emerged in the curve, in which the calculated energies are accurately converged to $E_0$ and not dependent on the value of $\tau$. When $\tau$ is greater than 0.75, $|\sin(E_0\tau)|$ is no longer the maximum, but $E(|\varphi_n\rangle)$ still converges to $E_0$. The main reason for this result is that the reference state is a Hartree-Fock (HF) state, which does not overlap with some of the lower order excited states. According to Eq. 1, even if $|\sin(E_1\tau)|$ is the maximum, due to $c_1 = 0$, the overlap of the first excited state with the $|\varphi_n\rangle$ state is always 0. In other words, even if the $|\sin(E_0\tau)|$ is not the maximum, but all of eigenstates with larger $|\sin(E_i\tau)|$ do not overlap with the reference states, so the ground state is still dominant in $|\varphi_n\rangle$. According to Figure 6(d), with $\tau$ is less than 0.62, the error becomes significantly large.

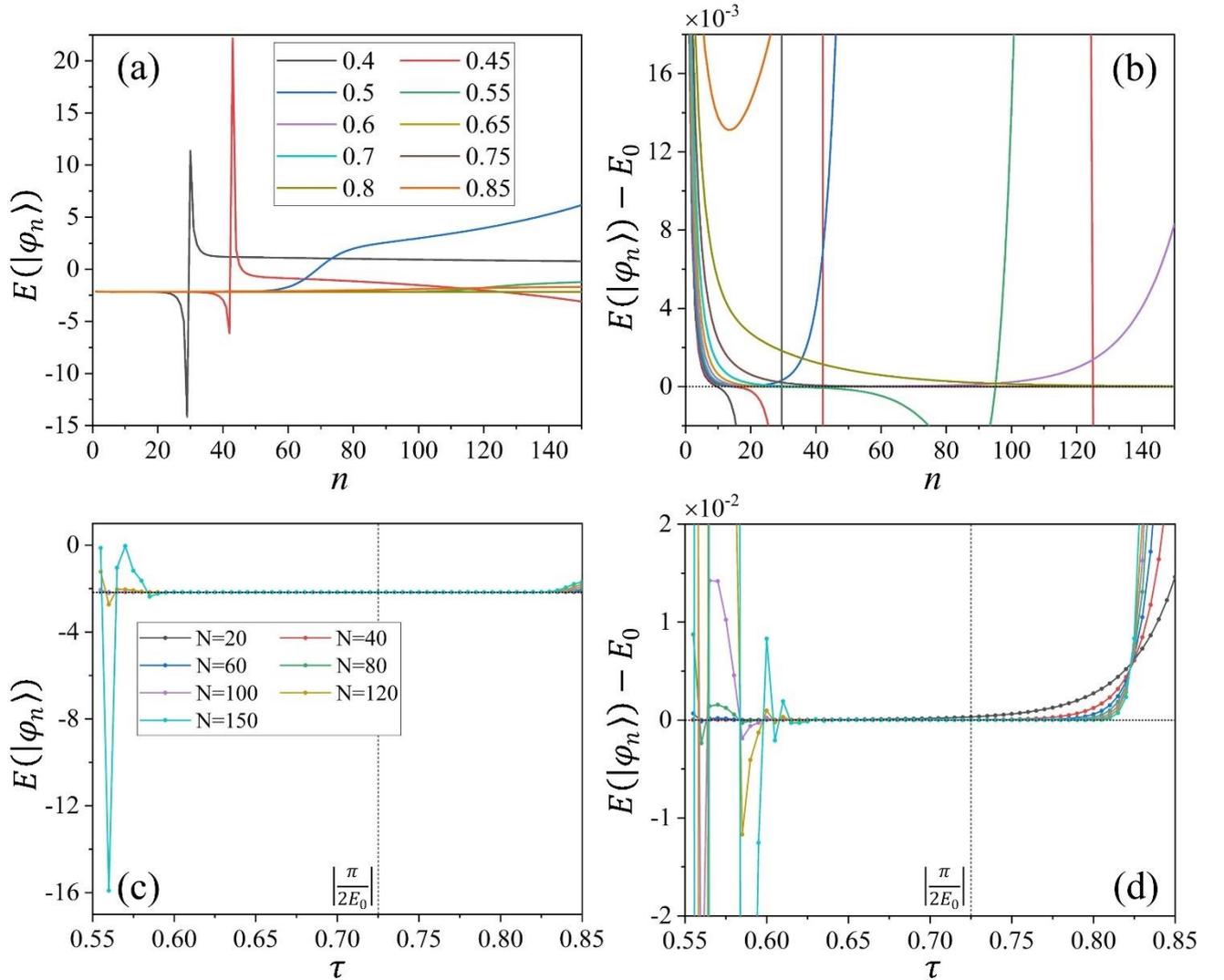

Figure 6. (a) Normalized energy (Y-axis, unit: hartree) of $|\varphi_n\rangle$ state ($E(|\varphi_n\rangle)$) as a function of power (X-axis) with different $\tau$ values. Different colors represent different $\tau$ values, which have been marked in the legend. (b) is enlarged view of (a) with the ordinate changes to $E(|\varphi_n\rangle) - E_0$. (c) Normalized energy (Y-axis, unit: hartree) of $|\varphi_n\rangle$ state ($E(|\varphi_n\rangle)$) as a function of $\tau$ (X-axis) with different powers. Different colors represent different powers, which have been marked in the legend. (d) is to (c) what (b) is to (a). Moreover, the horizontal dotted line is the position of $E_0$, and the vertical dotted line is the position of $\left|\frac{\pi}{2E_0}\right|$. All results are for the H$_4$ molecular chain with 1.0Å bond length (STO-3G basis set).

Figure 7 presents the error analysis for the normalization-coefficient based PSHO method described in Section 2.1. In Figure 7(a)(b), the Y-axis is $Q_n$ defined in Eq. 3. Similar with the energy based PSHO method, the curve appears abnormal oscillations when $\tau$ is 0.4 or 0.45. When $\tau$ is 0.5 or 0.55, the oscillation in the curve decays, but $Q_n$ still does not converge. $Q_n$ is converged when $\tau$ is in the range of 0.6 to 0.8. In order to further explore the errors associated with different $\tau$, Figure 7(c)(d) present $E'_n$ as a function of $\tau$, where the definition of $E'_n$ is in Eq. 4. In order to correctly converge to the ground state energy, the value of $\tau$ must be less than $\left|\frac{\pi}{2E_0}\right|$. When $\tau$ is between 0.62 and 0.72, the energy converges satisfactory, and a plateau is formed. The final converged energy is stable and not dependent on the value of $\tau$. Until $\tau$ is less than 0.62, the errors become large with the decreasing of $\tau$ and increasing of the power. These simulation results indicate that our analysis in Section 2.3 is correct. In order to suppress the error, the $\tau$ should be close to $\left|\frac{\pi}{2E_0}\right|$, and the final converged energy should not depend on the value of $\tau$.

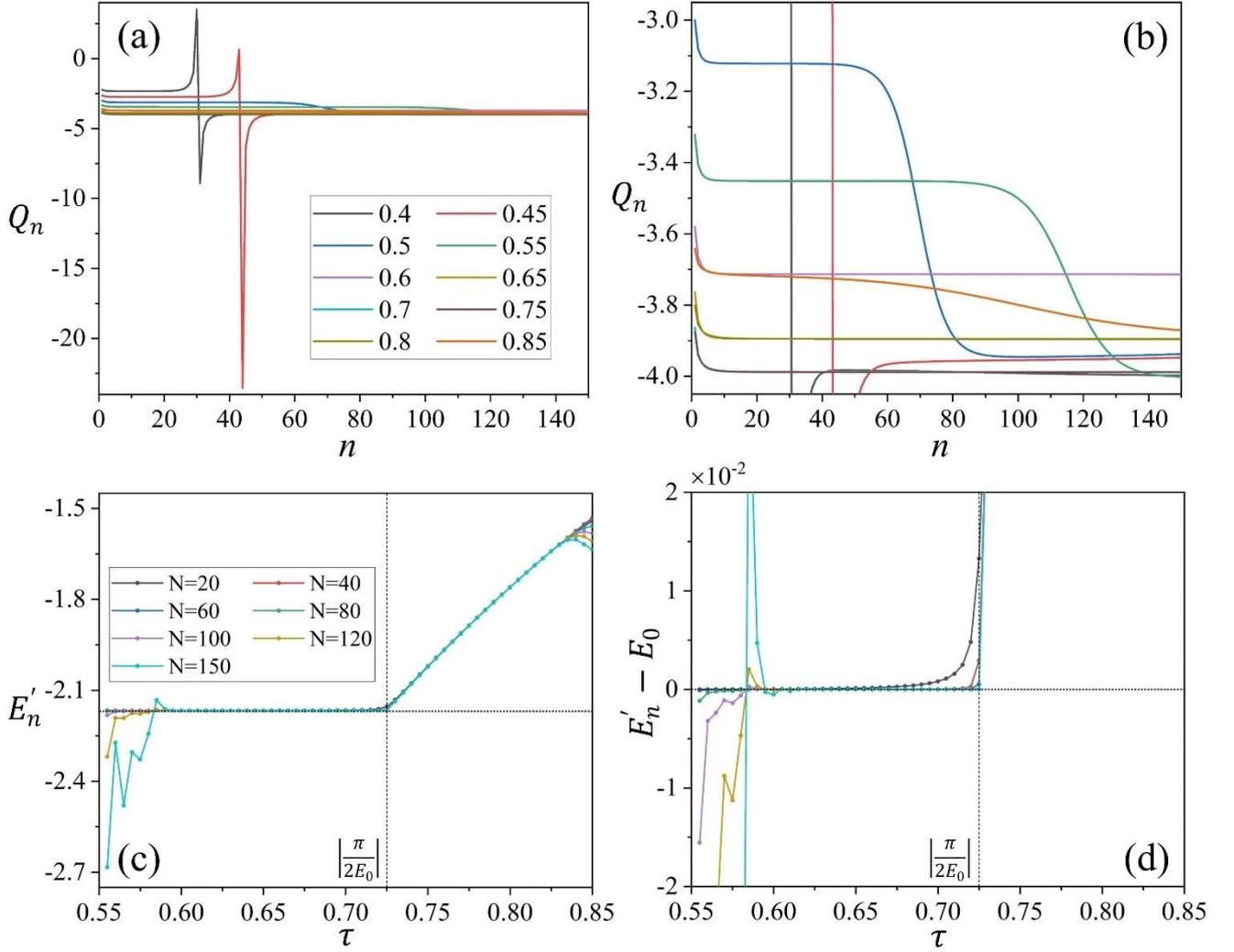

Figure 7. (a) $Q_n$ (Y-axis) as a function of power (X-axis) with different $\tau$ values. For the definition of $Q_n$, see Eq. 3. Different colors represent different $\tau$ values, which have been marked in the legend. (b) is enlarged view of (a). (c) $E'_n$ (Y-axis, unit: hartree) as a function of $\tau$ (X-axis) with different powers. Different colors represent different powers, which have been marked in the legend. (d) is enlarged view of (c) with the ordinate changes to $E'_n - E_0$. All results are for the H$_4$ molecular chain with 1.0Å bond length (STO-3G basis set).

### 3.3 Indirect PSHO Method for Ground State Calculation

According to the discussion in Section 2.3 and numerical results in Section 3.2, the $\tau$ value should be set close to $\left|\frac{\pi}{2E_0}\right|$ to suppress the errors in the indirect PSHO method. However, for an arbitrary molecular system, $E_0$ is unknown. The ground state energy can be determined by the following scheme. The initial reference state $|\varphi_0\rangle$ is set as HF state ($|\varphi_{\text{HF}}\rangle$), and set the initial $\tau$ as $\left|\frac{\pi}{2E_{\text{HF}}}\right|$. Since $|E_{\text{HF}}| < |E_0|$, $|E_0\tau_0| > \frac{\pi}{2}$, as can be seen in Figure 1(c), $|\sin(E_0\tau)|$ is generally not the maximum, the $\sin^n(\widehat{H}\tau)|\varphi_{\text{HF}}\rangle$ state may converge to other excited states, so the evaluated normalized energy is higher than $E_0$. Reducing $\tau$ gradually, and calculate the normalized energy (or normalization coefficient) of the $\sin^n(\hat{H}\tau)|\varphi_{\text{HF}}\rangle$ state with various $\tau$ by the indirect PSHO method. The ground state energy can be evaluated by Eq. 2 or Eq. 3. With $\tau$ decreases, the calculated energy will converge until the error appears. The converged energies can be regarded as the ground state energy. When the power is not large enough, the calculation results may deviate from the FCI results, but as the power increases, the energy will converge to the FCI energy.

For the $H_4$ molecule system, the converged energies with different powers are calculated and shown in Figure 8(a)(b). Compared with the energy-based PSHO method, the normalization-coefficient based PSHO method requires less sampling number, but its accuracy is inferior to that of the energy-based PSHO method. In particular, the energy-based PSHO method gives the energy close to FCI with the power greater than 20, whereas for the normalization-coefficient based PSHO method, the power should be greater than 30 to achieve the same accuracy. In order to further analyze the calculation errors, the energy difference ($\Delta E = E_{\text{PSHO}} - E_{\text{FCI}}$) is displayed in Figure 8(e). For the $H_4$ molecular chain system at equilibrium geometry, the HF state overlaps with the ground state significantly, i.e. $c_0$ is close to 1. In order to reach the chemical accuracy, the minimum power $n$ required in energy based and normalization-coefficient based PSHO methods are 10 and 20, respectively. With the bond stretched, $c_0$ is decreasing gradually, and the minimum power $n$ required to achieve chemical accuracy is increasing. For the energy based PSHO method, the power of 50 can ensure the calculated energies are within chemical accuracy for all bond lengths. However, for the same power in the normalization-coefficient based method, the errors exceed the chemical accuracy.

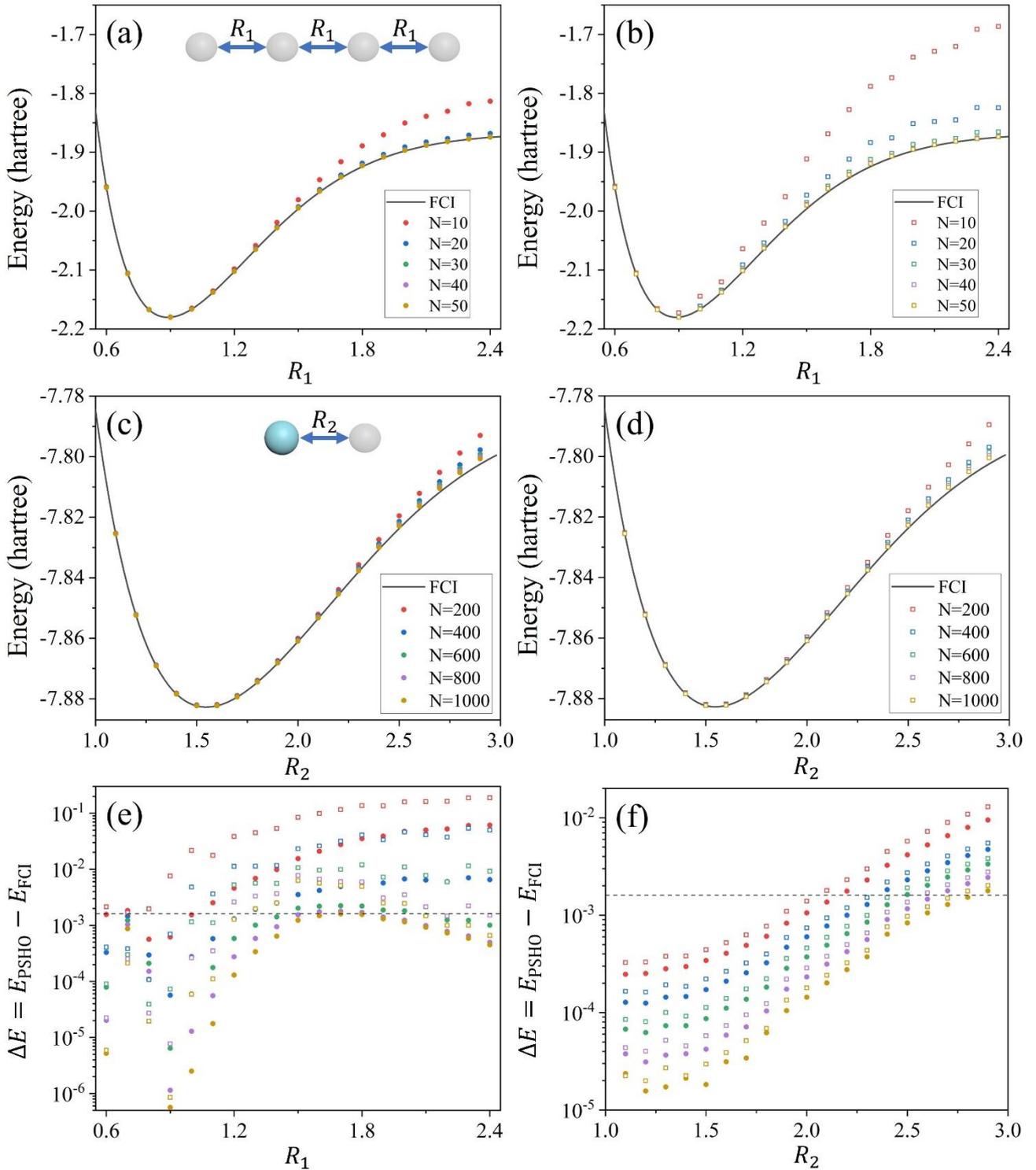

Figure 8. Simulation results for the ground state calculation of $H_4$(a)(b) and LiH(c)(d) molecule systems with different bond length, in which the lines are the results of the FCI method. The unit in Y-axis is hartree. Calculation results in (a)(c) are obtained by the energy based PSHO method, and results in (b)(d) are from the normalization-coefficient based PSHO method. Energy difference ($\Delta E = E_{\text{PSHO}} - E_{\text{FCI}}$) is also calculated for $H_4$(e) and LiH(f) systems, where the meanings of solid circle and hollow box are consistent with those in (a)(b)(c)(d). The horizontal dotted line is the line of chemical accuracy (0.0016 hartree).

For the LiH molecule, as can be seen in Figure 8(c)(d), the power required to achieve the chemical accuracy is much higher than that of $H_4$. This is mainly because the eigenvalues are close to each other, which causes the convergence to be slow. Take the ground state ($E_0$) and the first excited state ($E_1$) as an example. Their relative ratio is $\left|\frac{E_1}{E_0}\right|$, which is close to 1. According to the analysis in Section 2.1, in the $\sin^n(\widehat{H}\tau)|\varphi_0\rangle$ state, the relative proportion of the first excited state to the ground state is: $\frac{|c_1|^2 \sin^{2n}(E_1\tau)}{|c_0|^2 \sin^{2n}(E_0\tau)}$. The closer $\left|\frac{E_1}{E_0}\right|$ is to 1, the slower the relative proportion decreases, and a greater power is required to eliminate the first excited state. The corresponding calculation errors can be seen in Figure 8(f). When the LiH molecule is at the equilibrium geometry, a power of 100 for PSHO can reach the chemical accuracy. For the bond stretched systems, even the power of 1000 cannot reach the chemical accuracy. With the same power, the accuracy for the energy based PSHO method is higher than that of the normalization-coefficient based PSHO method.

In fact, the required power for reaching chemical accuracy can be reduced by offsetting a positive constant term in the Hamiltonian matrix. Take the ground state ($E_0$) and the first excited state ($E_1$) as an example. Suppose that the offset of the Hamiltonian matrix is $\varepsilon$, the ground state and the first excited state for the offset Hamiltonian is $E_0 + \varepsilon$ and $E_1 + \varepsilon$ respectively. Due to $E_0 < 0$ and $E_1 < 0$, the relative ratio $\left|\frac{E_1+\varepsilon}{E_0+\varepsilon}\right|$ deviates away from 1 with the increase of $\varepsilon$. Of course, $\varepsilon$ cannot be too large. For example, if $|E_1 + \varepsilon| > |E_0 + \varepsilon|$, then the $\sin^n(\widehat{H}\tau)|\varphi_0\rangle$ state no longer converges to the ground state. Therefore, assuming that the highest order eigenstate which has nonnegligible overlap with reference state is $|\Psi_i\rangle$, its corresponding energy is $E_i$. The offset should ensure $|E_0 + \varepsilon| > |E_i + \varepsilon|$. Given that $E_0 < 0$ and $E_i < 0$, the upper limit of $\varepsilon$ is $\left|\frac{E_0+E_i}{2}\right|$. Within this range, increasing $\varepsilon$ is beneficial to accelerate the convergence. Simulation results with $\varepsilon = 5.0$ can be seen in Figure 9. Compared the no-offset results in Figure 8(f), offset significantly improves the accuracy of calculation results. For the LiH at the bond length of 2.9Å, the minimum powers required to achieve chemical accuracy in the energy based and normalization-coefficient based PSHO methods are 600 and 1000, respectively. However, even for the energy based PSHO method with power is 1000, the result is still slightly exceeded the chemical accuracy with the no-offset Hamiltonian.

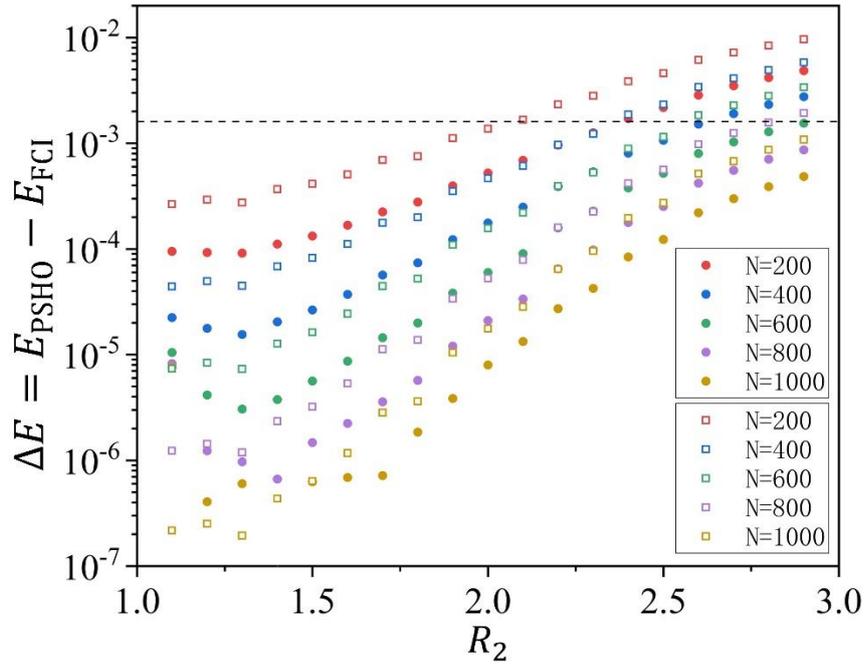

Figure 9. Simulation results for the LiH molecule with the offset is set to 5.0. Y-axis is the energy difference ($\Delta E = E_{\text{PSHO}} - E_{\text{FCI}}$). The legends of solid circle and hollow box in the figure are consistent with those in Figure 8.

### 3.4 Indirect PSHO Method for Excited States Calculation

For the excited states calculations, it is necessary to set the appropriate value of $\tau$ to converge to the targeted excited state. Similar with the ground state calculation, except that $\tau$ is gradually increase from $\left|\frac{\pi}{2E_0}\right|$. According to analysis in Section 2.1, when $\tau > \left|\frac{\pi}{2E_0}\right|$ or $|E_0\tau| > \frac{\pi}{2}$, the PSHO calculation will converge to the excited state that gives the maximum $|\sin(E_i\tau)|$ value. The energy of each excited state can be determined by the trend of the convergences for these curves. It is worth noting that some lower-order excited states may not overlap with the HF state. Therefore, it is necessary to use some other configurations as the reference states. Because the convergence of the energy based PSHO method is higher than that of the normalization-coefficient based PSHO method, only the energy based PSHO method is discussed in this section.

Figure 10 presents the results for the H$_4$ molecular chain system at the equilibrium geometry (bond length = 0.9 Å). Suppose that the ground state energy $E_0$ has been determined by the PSHO calculation, $\tau$ is changing from $\left|\frac{\pi}{2E_0}\right|$ to $\left|\frac{5\pi}{4E_0}\right|$. As shown in Figure 10, many plateaus are formed when $\tau$ increasing from $\left|\frac{\pi}{2E_0}\right|$ to $\left|\frac{5\pi}{4E_0}\right|$, and each plateau corresponds to the energy of an excited state. As the power increases, the plateaus in the $E(|\varphi_n\rangle)$-$\tau$ plots are more obvious. Different reference states produce different plateaus. For example, the first excited state cannot be obtained by the HF state, the reference state should be set to $|11011000\rangle$. In Figure 10(c), when $\tau$ is close to $\left|\frac{\pi}{E_0}\right|$, the curve tends to give a plateau when increasing the power. However, because the power is not large enough, the plateau is not obvious enough to obtain the corresponding eigenenergy. The curves in Figure 10(a)(c) show a rough monotonicity, that is, the normalized energy increases monotonically with $\tau$, and so is the energy of each plateau. This is consistent with the analysis in Figure 1(c). With the increase of $\tau$, the final state will converge to each excited state in order. However, an anomaly appears in Figure 10(b). When $\tau$ is near $\left|\frac{\pi}{E_0}\right|$, the curve rises sharply to positive and then falls back with the increase of $\tau$. This is mainly because the PSHO method can only filter out the states with the largest absolute value of $\sin(E_i\tau)$. If there is an eigenstate with positive energy in the reference state, the final state can converge to this state when $\tau$ takes a certain value.

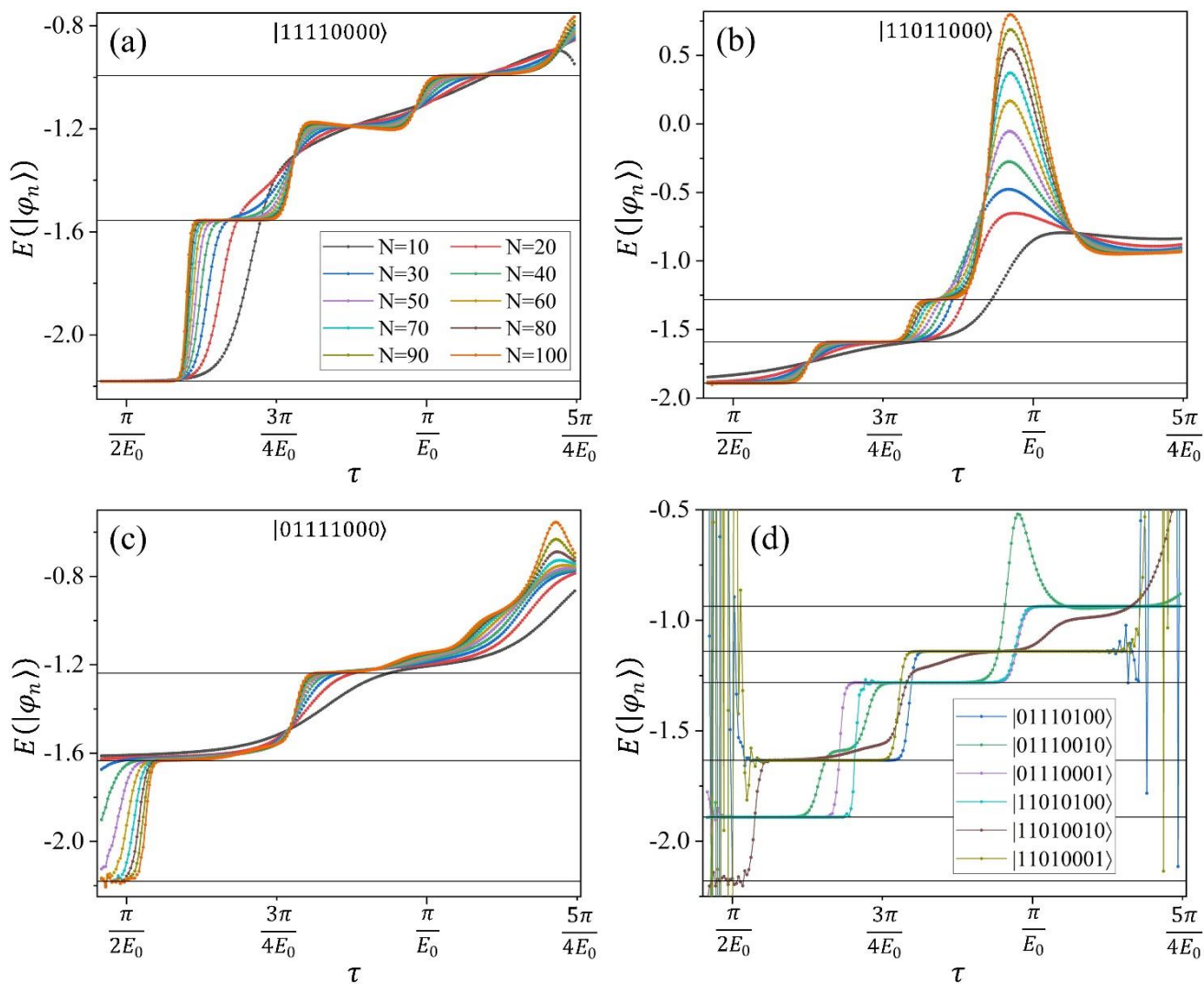

Figure 10. (a)(b)(c) Normalized energy (Y-axis, unit: hartree) of $|\varphi_n\rangle$ state as a function of $\tau$ (X-axis) with different powers. Different colors represent different powers, which have been marked in (a) legend. The difference in (a)(b)(c) figures is that different reference states are used, where the HF state ($|11110000\rangle$) is used in (a), $|11011000\rangle$ state is used in (b) and $|01111000\rangle$ state is used in (c). (d) Normalized energy of $|\varphi_{100}\rangle$ state as a function of $\tau$ (X-axis) with different reference states. The result is for the H$_4$ molecular chain system at equilibrium geometry (0.9Å

bond length). The horizontal solid lines in these figures are eigenvalues of the system calculated by FCI method.

It is worth noting that there seems to show a platform around $\tau = 0.875 \left|\frac{\pi}{E_0}\right|$ (the ordinate is around -1.2) in Figure 10(a). If we look Figure 10(a) closely, we find that this is not a plateau. The real platform is that the energy-τ curve tends to be horizontal with the increase of power. However, around -1.2, with the increasing power, the curve first becomes horizontal, and then becomes steep in the opposite slope direction, and does not converge to a plateau. The reason for this phenomenon may be due to the fact that two eigenstates are on both sides of the curve (around $\tau = 0.8 \left|\frac{\pi}{E_0}\right|$ and $\tau = 0.95 \left|\frac{\pi}{E_0}\right|$). When the power is large enough, platforms will appear on both sides, and the previous trick plateau (around $\tau = 0.875 \left|\frac{\pi}{E_0}\right|$) evolves into a slope line connecting the two real platforms. Strictly speaking, the normalized energy in Figure 10(a) curve does not increase monotonically with $\tau$. The reason is same as that in Figure 10(b). In Figure 10(a), the power is not high so that the curve non-monotonic tendency is not obvious. In Figure 10(d), the power is set as 100, and many single excited configurations are used as reference states. Some curves show the abnormal oscillations in certain ranges of $\tau$. It is worth noting that the energies of the plateaus perfectly match with the energies of the excited states.

As mentioned above, the energies of excited states can be obtained by taking the values of the plateaus with different reference states. However, the number of configurations is factorially dependent on the number of spin orbitals, and it is impractical and unnecessary to using all configurations as the reference states. In general, the overlap between the lower-excitation configurations and the lower-order eigenstates is nonnegligible. In the present study, only the HF state and all of the single excited configurations are considered. As can be seen in Figure 11, all PSHO energies for excited states are in good agreement with the FCI results. The majority of eigenenergies can be calculated with the PHSO method, but there are still a few missing eigenstates. This is mainly due to the fact that these missing states do not overlap with the employed reference states. To obtain these states, the reference state can be extended to configurations of double or higher excitations.

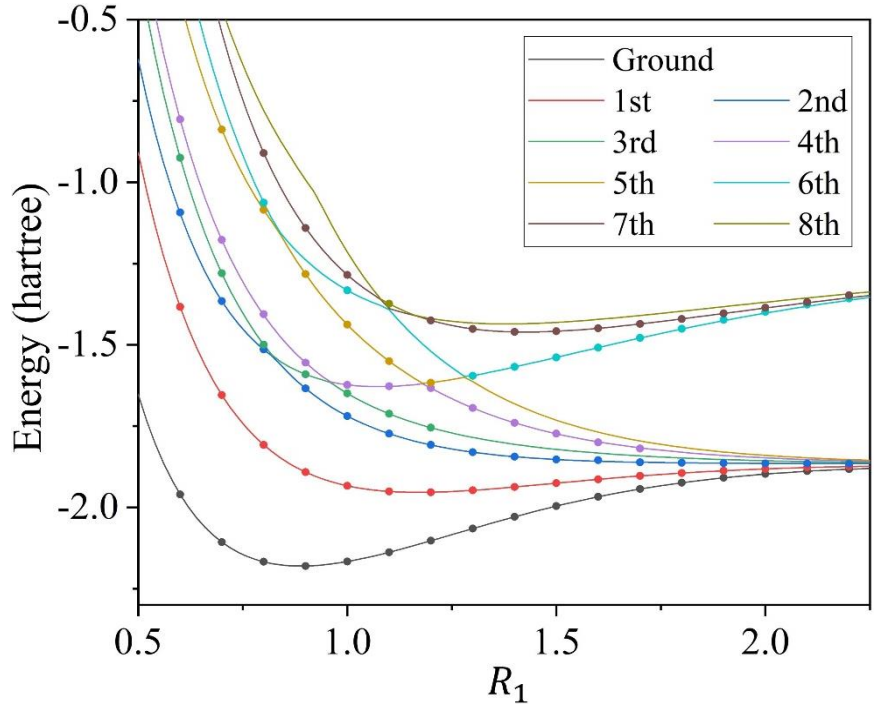

Figure 11. Ground and excited-state energies of $H_4$ molecule systems with different bond length. The lines are the results of the FCI method, and the dots are from the PSHO method. Different colors represent different eigenvalues, which is marked in the legend.

**3.5 Feasibility Discussion**

The purpose of developing quantum computing algorithm is to transform classical computing problems with exponential complexity into the tasks of polynomial complexity. The eigenstate calculations of a molecular system are of exponential complexity on classical computers. In the PSHO method, the number of qubits increases linearly with the size of the molecular system. The circuit depth and measurement number are related to the power. The power is linearly related to the time of the $e^{\pm i\hat{H}\tau}$ operator. According to Eq. 19, The circuit depth is linearly related to the square root of power, and the measurement number is linearly related to the power, as shown in Eq. 10 and Eq. 11. Therefore, in order to analyze the computational complexity of the PSHO method, it is necessary to determine the minimum power to ensure the convergence of a PSHO calculation.

In the case of $|E_0\tau| < \frac{\pi}{2}$, the relative fraction of all excited states in $|\varphi_n\rangle$ is exponentially decayed with the power $n$ increasing, but this exponential decreasing rate is not as rapid as $2^{-n}$. Here we are concerned about the "relative fraction". According to the discussion in Section 3.3, the closer $\left|\frac{E_1}{E_0}\right|$ is to 1, the slower the relative proportion decreases, and a greater power is required to eliminate the first excited state. To alleviate the high power problem, in the last paragraph of section 3.3, we proposed that the convergent power can be reduced by offsetting a positive constant term in the Hamiltonian. The energy-shift can amplify relative gaps between different eigenstates. However, subjected to the highest order eigenstate with non-zero overlap with reference state, the energy-shifting must meet a specific requirement. Therefore, the energy-shifting method has limited effect for the power reduction. Here are a couple of questions we need to address for the PSHO method: what is the minimum power required for the convergence with the increased system size? Whether it is increasing exponentially with the system size? The energy gaps between different eigenstates are not directly related to the size of the system, so it is difficult to determine how the power scales with respect to the size of the system. However, we can analyze the problem from other perspective. For the Hamiltonian concerned in quantum chemistry, its corresponding Hilbert space is exponentially scale with the system size. However, the state we care about only covers a minimal subspace in the Hilbert space. Take the ground state as an example. Generally speaking, most of the configurations are not involved in the ground state. Therefore, it is reasonable to believe that, eigenstates which have non-zero overlap with a specific reference state is also in a minimal subset of the whole eigenstate set. In the PSHO method, taking appropriate energy shift and $\tau$ value to make all $E_i\tau$ values are in the interval of $\left[-\frac{\pi}{2}, 0\right]$. The number of states in this interval will not scale exponentially with the system size, and the average gap between states will not decrease exponentially. For any two eigenstates, their $\sin(\hat{H}\tau)$ operator eigenvalues are not infinitely close. Therefore, we believe that the minimum power will not increase exponentially with the increase of system size.

The above analysis assumes that all eigenvalues are evenly distributed, in which energy gaps of each adjacent eigenstates are equal. However, we know that this not true for the real system. The convergence rate of PSHO depend on the energy spectrum distribution. For the energy spectrum of hydrogen atom, we know that the lower the energy level, the greater the energy gap, and the higher the energy level, the smaller the energy gap. We believe that similar trend also holds in molecular systems. Generally, we only focus on some eigenstates at low energy levels, so the minimum power required for calculating eigenstates in PSHO method may be lower than the above analysis for the evenly distributed eigenvalues.

For the ground state calculation, $\tau$ has to be chosen such that $|E_0\tau| < \frac{\pi}{2}$, which means that $\tau$ will be very small for large energies. Too small $\tau$ may cause some difficulties for the implementation of PSHO on the actual quantum devices. There are two methods to address this issue. The first one is the energy-shifting method mentioned in the last paragraph section 3.3. Giving that the ground state energy is generally negative in quantum chemistry, the positive energy-shifting on the Hamiltonian makes the shifted ground state energy closer to 0, so that the value of $\tau$ does not have to be too small. The second method is changing the $|E_0\tau| < \frac{\pi}{2}$ constraint to $|E_0\tau| < \frac{\pi}{2} + m\pi$, where $m = 0,1,2\cdots$. When the $|E_0|$ is large, too small $\tau$ can be prevented by increasing the value of $m$. However, increasing the $m$ may cause the PSHO state converge to other excited states, instead of the ground state. When $|\sin(E_i\tau)|$ of an excited state is closer to 1 than $|\sin(E_0\tau)|$, the PSHO state will converge to the $E_i$ state.

From the above analysis, we can see that the PSHO method might not be applicable to all Hamiltonians. Based on some assumptions, we only theoretically demonstrate that the convergence power does not increase exponentially with the size of the system. It is difficult to derive the exact growth trend of power. However, we still believe that the PSHO method has advantages in solving some systems that are difficult to calculate by other methods. Compared with the popular VQE algorithm, PSHO does not require a parameterized ansatz circuit, and the complex nonlinear optimizations are avoided. In VQE, ansatz circuit construction and parameter optimization greatly affect the result accuracy. Moreover, the measurement statistical error in VQE affects not only the energy estimation but also the gradient determination, so the PSHO method is more robust than VQE. In terms of quantum resource cost, the number of qubits in PSHO is only one more than that of VQE. For the circuit depth, it is also an open question how deep an ansatz needs to be in VQE to represent the ground state. This often depends on the specific system. Thus, it is difficult to compare their circuit depth. For the measurement numbers, both VQE and PSHO method need to calculate the expectation value of Hamiltonian, and their measurement numbers are the same. However, VQE needs to estimate energy and all parameter gradients and perform it multiple times because of the iterative optimizations. The circuit in Figure 3 should be implemented $n$ (power) times in PSHO, and the measurement number will increase $n$ times. It is difficult to determine the number of parameters and iterations. Thus, it is also difficult to compare their measurement numbers.

It needs to be clarified that although PSHO method has some theoretical advantages over VQE, the purpose of developing the PSHO method is not to replace VQE method, but to be a complementary method with VQE. As in many classical quantum chemical methods, each method has its own pros and cons. For some systems, such as LiH at the equilibrium geometry, the VQE method can easily obtain the accurate ground state energy. However, for some non-equilibrium systems or excited states calculations, it may be difficult to construct the ansatz that can effectively prepare the eigenstate. For these cases, the PSHO method may be a useful alternative. If the VQE ansatz circuits are too complex to be optimized, one can also try the PSHO method. Moreover, it is also a good idea to combine VQE with PSHO method. As we mentioned in Section 2.2, using VQE state as the reference state in the PSHO algorithm, the power required to converge to the ground state may be less than that of the HF state. These are worthy of further studies in the future.

## 4. Conclusions

In this paper, we introduce a new hybrid classical-quantum algorithm, named as power of sine Hamiltonian operator (PSHO), to evaluate the normalized energy of the $\sin^n(\widehat{H}\tau)|\varphi_0\rangle$ state. By increasing the power $n$, the normalized $\sin^n(\widehat{H}\tau)|\varphi_0\rangle$ state will converge to the $|\Psi_i\rangle$ state with the maximum $|\sin(E_i\tau)|$ value in $|\varphi_0\rangle$. The PSHO method can be used to determine eigenvalues of a given Hamiltonian ($\widehat{H}$). PSHO can be implemented via a direct or indirect algorithm. The direct algorithm attempts to prepare the normalized $\sin^n(\widehat{H}\tau)|\varphi_0\rangle$ states on a quantum register. Since the probability of successfully preparing the target state decreases exponentially with the increase of power, the computational cost of the direct algorithm is prohibitively high. Nevertheless, the quantum circuit in the direct method lays a solid foundation for the indirect algorithm. The indirect algorithm can calculate the normalized energy without prepare the corresponding quantum state, where each paired term in its algebraic expansion can be evaluated independently. It is necessary to suppress the errors in the indirect PSHO method. For the ground state calculation, the $\tau$ should be close to $\left|\frac{\pi}{2E_0}\right|$. Since $E_0$ is unknown, the initial $\tau$ value can be set to $\left|\frac{\pi}{2E_{\text{HF}}}\right|$. Gradually reduce the value of $\tau$ and increase the power until the evaluated energy converges. For the excited states calculation, the $\tau$ should be gradually increased from $\left|\frac{\pi}{2E_0}\right|$. The energy of each plateau in the energy-$\tau$ plot corresponds to an the energy of an eigenstate. In addition, even only the normalization coefficient of the $\sin^n(\widehat{H}\tau)|\varphi_0\rangle$ state is obtained, the eigenenergies can be extracted from the normalization coefficients. Compared with the energy based PSHO method, the normalization-coefficient based PSHO method requires less sampling number, but its accuracy is inferior to that of the energy based PSHO method. The performance of PSHO is demonstrated by numerical simulations. For the $H_4$ molecular chain, the PSHO calculated energies agree well with those of FCI. For the LiH molecule, the power required to achieve chemical accuracy is much larger that for $H_4$, but it can be highly reduced by offsetting the Hamiltonian. It is difficult to derive how the minimum power required for convergence scales with respect to the size of the system. But we theoretically demonstrate that the power does not increase exponentially with the system size based on some assumptions. It is difficult to compare quantum resource cost of VQE and PSHO, and PSHO does not need the design of ansatz circuits and the complex nonlinear optimizations. Compared with nonunitary imaginary time evolution method, the indirect PSHO method avoids the difficulty that nonunitary operators cannot be directly transformed into quantum circuits. This work extends the quantum algorithms for the eigenstates calculations and broadens the applications of quantum computing in quantum chemistry.